\documentclass[conference]{IEEEtran}
\IEEEoverridecommandlockouts

\usepackage{cite}
\usepackage{amsmath,amssymb,amsfonts}

\usepackage{graphicx}
\usepackage{textcomp}
\usepackage{xcolor}
\def\BibTeX{{\rm B\kern-.05em{\sc i\kern-.025em b}\kern-.08em
    T\kern-.1667em\lower.7ex\hbox{E}\kern-.125emX}}

\usepackage[utf8]{inputenc} %
\usepackage[T1]{fontenc}    %
\usepackage{hyperref}       %
\hypersetup{hidelinks}
\usepackage{xurl}

\usepackage{url}            %
\usepackage{booktabs}       %
\usepackage{amsfonts}       %
\usepackage{nicefrac}       %
\usepackage{microtype}      %
\usepackage{xcolor}         %
\usepackage{amsmath}
\usepackage{graphicx}
\usepackage{float}
\usepackage{multirow}
\usepackage{enumitem}

\usepackage[caption=false,font=footnotesize]{subfig}
\usepackage{calc}
\usepackage{tabularx,array}

\usepackage{algorithm}      %
\usepackage{algpseudocode}  %
\usepackage{amssymb}        %

\usepackage{pifont, makecell, multirow, nicematrix}
\usepackage[table]{xcolor}
\newcommand{\cmark}{\ding{51}}  \newcommand{\xmark}{\ding{55}}  \newcommand{\pmark}{$\circ$}
\colorlet{shade}{gray!10}

\usepackage{tikz}
\newcommand{\circled}[1]{%
  \tikz[baseline=(char.base)]{%
    \node[shape=circle, draw,
          inner sep=0.65pt, line width=0.5pt,
          font=\small, minimum size=1.6ex,
          yshift=0.5ex] (char) {\raisebox{+0.2ex}{#1}};%
  }%
}%

\newif\ifanon
\anontrue 
\begin{document}

\title{Dissecting the Carbon--SLO Tension in Geo-Distributed LLM Serving\\
}

\author{
\IEEEauthorblockN{Burak Topcu, Musa Cim, Zuhal Ozturk, Mahmut Taylan Kandemir}
\IEEEauthorblockA{\textit{Department of Computer Science and Engineering}\\
\textit{The Pennsylvania State University}\\
State College, PA 16801, USA\\
\{bvt5283, mtc5693, zbo5031, mtk2\}@psu.edu}
}
\pagestyle{plain}

\maketitle

\begin{abstract}
Large Language Model (LLM) inference is driving a growing share of datacenter carbon emissions. Although carbon-aware scheduling can reduce these emissions in geo-distributed datacenters by exploiting the \textit{spatiotemporal} variations in grid carbon intensity, the latency-sensitive nature of LLM serving constrains this design space. While prior studies investigate the carbon savings achievable through spatial scheduling, they oversimplify the dynamics of modern inference engines, such as queueing and continuous batching, which strongly shape inference latency. We address this issue by designing and evaluating a latency-centric, geo-distributed inference simulator, calibrated with hardware-grounded latency and power measurements, that reproduces these dynamics and enables the study of carbon savings under strict service-level objectives (SLOs). We then formulate a series of SLO- and carbon-aware schedulers of increasing sophistication, from per-request heuristics to batch-level optimization: (i) a greedy heuristic, (ii) an integer linear programming (ILP)-based scheduler jointly assigning batches of requests, (iii) an iterative ILP-based solution projecting its own assignments back into the load estimates to capture load coupling, and (iv) a facility mixed-integer linear programming (MILP) scheduler pricing the energy of batched requests collectively rather than in isolation. Across three LLMs under MLPerf SLOs and varying system loads, the MILP-based option offers the best carbon--SLO balance, reducing carbon emissions by up to 60\% relative to the round-robin policy, with an average SLO violation of 3\%. Our characterization further shows that savings grow with the spread of carbon intensity across datacenters and shrink as system load increases, while SLO attainment depends on how frequently schedulers receive datacenters' load updates.
\end{abstract}


\section{Introduction} \label{sec:intro}
Over the past decade, LLM-powered applications have become widespread across domains~\cite{ai_app_1, ai_app_2, AI_IndexRep}, driven by breakthroughs in reasoning~\cite{CoT, SelfCons_CoT} and agentic pipelines~\cite{ReAct, AutoGen, CodeAgents}, turning inference into a continuously running workload worldwide. Sustaining this inference demand has spurred massive investments in accelerator-rich datacenters~\cite{mckinsey_invest, doe_invest1, doe_invest2} and raised concerns about their environmental impacts~\cite{concern, concern_2}. The International Energy Agency projects the energy use of accelerator-equipped servers to grow by 30\% annually, over three times faster than conventional ones, and to reach nearly half of net datacenter energy use through 2030~\cite{iea2025energyAI}. Beyond energy consumption, this growth directly translates into carbon emissions through the carbon intensity (hereafter, CI) of the grids powering these sites, increasing their share of the overall carbon chain~\cite{ict_co2_share}. Recent studies~\cite{reducingCarbGenAI, carbonLifeCycleLLM} show that inference already emits over 25$\times$ more carbon than training throughout a model's lifetime, underscoring the need for carbon-efficient inference serving.

Despite the adoption of green energy-based solutions~\cite{Net0_AWS, Net0_google}, renewable availability varies \textit{temporally} and \textit{spatially}~\cite{tempLetsWait, carbonCast}, limiting their capacity to address inference-related emissions. Complementing these strategies, prior work has traded either output quality or serving performance for carbon reduction. The former selects the carbon-optimal LLM that meets an accuracy target~\cite{gar, clover} or shortens outputs through prompt-level directives~\cite{sprout}. The latter preserves output quality while trading latency for lower emissions, either by reducing energy use or by scheduling inference to greener datacenters. As energy efficiency has been studied across the compute stack for decades~\cite{CompArch_PowEff, EnEff_Sw, EnAwareSch_DCs}, recent efforts exploit the remaining slack through frequency scaling under latency service-level objectives (SLOs)~\cite{throttLLM, uServe} or load-adaptive inference configurations~\cite{dynamoLLM}. However, carbon savings with these energy-oriented approaches diminish as the load increases, limiting their effectiveness at scale. In contrast, CI varies by more than an order of magnitude across regions and over time~\cite{elecMaps}, offering significant potential to reduce inference-related emissions through carbon-aware scheduling while preserving output quality.

Temporal scheduling that suspends and resumes jobs based on green energy availability~\cite{tempLetsWait} is appropriate for delay-tolerant workloads but not for latency-critical ones, such as LLM inference with stringent time-to-first-token (TTFT) and time-per-output-token (TPOT) objectives. Spatial scheduling can meet latency constraints, subject to data-placement regulations such as GDPR~\cite{gdpr}, but suffers from load imbalance, as carbon-greedy policies dispatch requests to the greenest datacenters and inflate per-token latency. While carbon-aware \textit{spatiotemporal} scheduling has been studied for batch ML training, HPC~\cite{ecovisor, limitationsPaper, carbonScaler}, and even for latency-sensitive web services~\cite{casper, carbonClipper}, only a small but growing body of work explores its \textit{spatial} axis for LLM inference across geo-distributed datacenters.

\begin{table*}[ht]
\centering
\small 
\setlength{\tabcolsep}{3pt}
\renewcommand{\arraystretch}{0.95}
\setlength{\aboverulesep}{0pt}
\setlength{\belowrulesep}{1pt}
\definecolor{hdrA}{gray}{1}
\definecolor{hdrB}{gray}{1}
\definecolor{oursrow}{RGB}{226,238,252}
\caption{Carbon-aware geo-distributed LLM scheduling approaches (\cmark{} modeled, \xmark{} not modeled, \pmark{} partial, -- not applicable).}
\vspace{-1mm}
\label{tab:related}
\begin{NiceTabular}{@{}l l l cc cc l@{}}[code-before =
  \rectanglecolor{hdrA}{1-1}{2-1}
  \rectanglecolor{hdrB}{1-2}{2-3}
  \rectanglecolor{hdrA}{1-4}{2-5}
  \rectanglecolor{hdrB}{1-6}{2-7}
  \rectanglecolor{hdrA}{1-8}{2-8}
  \rowcolor{oursrow}{9}]
\toprule
\multirow{2}{*}[-0.6ex]{\textbf{Study}} & \multicolumn{2}{c}{\textbf{(i) Formulation}} & \multicolumn{2}{c}{\textbf{(ii) Batch-aware}} & \multicolumn{2}{c}{\textbf{(iii) Overheads}} & \multirow{2}{*}[-0.6ex]{\textbf{Reported latency}} \\
\cmidrule(lr){2-3} \cmidrule(lr){4-5} \cmidrule(lr){6-7}
 & Placement & Latency target & Latency & Power & Queue & Transfer & \\
\midrule
SLIT~\cite{slit} (2025)           & Fixed 15-min plan & Min.\ TTFT, unbounded           & \xmark & \xmark & \xmark & \cmark & Only mean TTFT \\
Green-LLM~\cite{greenllm} (2025)  & Fixed 1-h plan    & Mean E2E $\le$ 5\,s             & \xmark & \xmark & \xmark & \cmark & None (\$ penalty) \\
FREESH~\cite{freesh} (2025)       & Fixed 30-min plan & E2E $\le 5\times$ profiled      & \xmark & \xmark & \xmark & \xmark & E2E viol., p99 \\
GCOS~\cite{gcos} (2026)           & Per job (GPU-h)   & Job wait $\le$ 10--20\% runtime & --     & \xmark & \xmark & \cmark & Wait violation \\
MARLIN~\cite{marlin} (2026)       & Fixed 15-min plan & Min.\ TTFT, unbounded           & \xmark & \xmark & \xmark & \cmark & Only mean TTFT \\
MOER~\cite{moerRouting} (2026)    & Per request       & None                            & \xmark & \pmark & \xmark & \xmark & E2E p95 \\
\midrule
\textbf{Ours}                     & Per request and batch & TTFT, TPOT $\le$ MLPerf & \cmark & \cmark & \cmark & \cmark & SLO att., p\{50,90,99\} \\
\bottomrule
\end{NiceTabular}
\vspace{-4mm}
\end{table*}

Table~\ref{tab:related} summarizes studies that aim to reduce inference-related emissions by scheduling requests across datacenters using linear or integer programs~\cite{greenllm, freesh}, metaheuristics~\cite{slit}, or learning-based search~\cite{marlin, gcos}. Unfortunately, these studies overlook several key facets of real-world inference scenarios: (i) most~\cite{slit, greenllm, freesh, marlin} schedule requests to datacenters at coarse time intervals and assume load-independent latency, which hinders capturing how TTFT and TPOT change in practice as continuous batching~\cite{orca} dynamically admits and completes requests; (ii) none fully models power as a function of inference workload and phase, although \textit{prefill} and \textit{decode} exhibit distinct compute utilization~\cite{splitWise}, driven by the prefill length and the batch size, respectively; (iii) none accounts for the queueing delay when an inference engine lacks memory for a new request, and some~\cite{freesh, moerRouting} further ignore inter-region transfer delays; and (iv) their latency targets are mostly self-defined rather than grounded in standard benchmarks such as MLPerf~\cite{mlperf}. Consequently, these studies oversimplify practical latency and power dynamics and lack representative evaluation against benchmark-grounded per-token SLOs, limiting their applicability to real-world deployments.

To study carbon-aware LLM scheduling under real-world serving conditions, we develop a latency-centric, geo-distributed serving simulator.\footnote{\url{https://github.com/topcuburak/carbon-slo-at-geoscale}} It instantiates datacenters across multiple regions, each with GPU-equipped nodes hosting configurable data-parallel inference engines and forwards requests between regions~\cite{slit, freesh, marlin} over measured inter-region network latencies~\cite{cloudping}. Engines capture iterative continuous batching behavior~\cite{orca}, model queueing delays by admitting requests only when sufficient memory is available, and estimate per-inference phase latency and power using analytical models calibrated across different models, degrees of parallelism, and datasets. Using this simulator, we further design and test a suite of carbon-aware schedulers, from a lightweight carbon-weighted greedy policy to integer programs, and characterize their trade-offs among carbon, SLO attainment, and tail latency against both carbon-blind and carbon-aware baselines~\cite{greenllm, moerRouting}. Our novelty lies in identifying the engine-level datacenter state, namely the active batch load and queue depth, from which a carbon-aware scheduler can estimate TTFT and TPOT at scheduling time, which, to our knowledge, no prior carbon-aware geo-distributed scheduler has explored. Our main {\bf contributions} are as follows: 

\begin{itemize}[topsep=1pt, itemsep=0.5pt, parsep=0pt, leftmargin=1em]
    \item \textbf{High-fidelity simulator:} An extensible simulation framework for geo-distributed LLM serving, validated end-to-end within a node against real inference, that captures inter-region transfers and key serving features, such as continuous batching and engine queueing, using calibrated per-phase latency and power models over configurable topologies and scheduling policies.
    \item \textbf{SLO- and carbon-aware schedulers:} A series of schedulers of increasing sophistication, from per-request heuristics to batch-level optimization:
    \begin{itemize}[leftmargin=0.6em, labelwidth=0.6em, labelsep=0.3em, align=left, topsep=0pt, itemsep=0pt]
        \item \textbf{Greedy:} minimizes a weighted carbon-latency score per request at negligible overhead.
        \item \textbf{Standard integer linear program (ILP):} jointly assigns batches of requests, treating each datacenter's load as fixed during the assignment. 
        \item \textbf{Iterative ILP:} re-projects datacenter batch states after each ILP solve to capture load coupling and re-solves until the assignment converges or an iteration limit is reached.
        \item \textbf{Facility mixed-integer linear program (MILP):} models batched energy as a shared per-datacenter cost and a marginal per-request cost, discouraging the scattering of a batch across many datacenters.
    \end{itemize}
    \item \textbf{Carbon--SLO frontier characterization:} A comprehensive evaluation of carbon savings and SLO attainment across models, prompt lengths, system loads, and CI variation against carbon-blind and carbon-aware baselines. The facility MILP achieves the best carbon–SLO balance, reducing carbon emissions by up to 60\% relative to the round-robin policy with an average SLO violation of 3\%, whereas the most carbon-aggressive greedy setting, at comparable savings, violates the SLO for up to 55\% of requests. We further analyze the TTFT tail behavior, scheduling overhead, sensitivity to load-update frequency, topology, robustness to facility overheads, and predicted output lengths.
\end{itemize}

\section{Background} \label{sec:background} %

\noindent {\bf Carbon intensity (CI) and carbon accounting.} Electricity grids are supplied by a mix of sources $\mathrm{G}$, each with a distinct CI in gCO$_2$/kWh. The grid's overall CI is a generation-weighted average \scalebox{0.9}{$CI = \sum_{g \in \mathrm{G}} (P_g / P_{\mathrm{total}}) \cdot CI_g$}, where $P_g$ and $CI_g$ denote the power contribution and intensity of source~$g$, and $P_{\mathrm{total}}$ is the total grid power. Low-carbon sources such as nuclear and solar have median CIs below 50, whereas fossil sources such as natural gas and coal have medians of 490 and 820, respectively~\cite{ipcc_ci}. Fluctuating renewable availability and power demand on grids over time change the energy-source mix, leading to a \textit{spatially} and \textit{temporally} varying CI. 

The Greenhouse Gas (GHG) Protocol~\cite{GHGScopes} is an international standard for measuring, managing, and reporting greenhouse gas emissions from operations, value chains, and mitigation actions. It defines three emission scopes, with emissions from powering datacenters (hereafter, DCs) used for LLM inference falling under Scope 2, also referred to as \textit{operational emissions}~\cite{chasingCarbon}. CI signals are sampled at fixed intervals (e.g., hourly), and the operational emissions of a DC over $K$ intervals are quantified as \scalebox{0.9}{$\sum_{k=1}^{K} E_{\mathrm{tot}}[k] \cdot CI[k] $}, where \scalebox{0.9}{$E_{\mathrm{tot}}[k] = P_{\mathrm{tot}}[k] \cdot \Delta t$} is the total energy consumption in the interval $k$ while operating with a total average power \scalebox{0.9}{$P_{\mathrm{tot}}[k]$}. The DC power is \scalebox{0.95}{$P_{\mathrm{tot}} = P_{\mathrm{IT}} \times \mathrm{PUE}$}, where \textit{power usage effectiveness} (PUE) accounts for facility overheads such as cooling and lighting.

\noindent {\bf Geo-distributed LLM serving.} Online (i.e., server) or interactive LLM inference scenarios~\cite{mlperf} are evaluated under per-token latency objectives: TTFT, which covers queueing and prefill durations, and TPOT, which corresponds to the average time for output token generation~\cite{distServe}. Both latency metrics depend on the engine's load and its serving features. For example, continuous batching~\cite{orca} can admit new requests at each iteration, shortening queueing and hence TTFT; however, each admitted prefill intervenes in the ongoing decode pipeline and extends the TPOT of active requests. Chunked prefill~\cite{sarathiServe} and disaggregated serving~\cite{distServe, splitWise} mitigate this by splitting prefills across steps or running prefill and decode on separate resources, offering control over the TTFT--TPOT balance.

To meet these latency SLOs under varying system loads at scale, deployments place data-parallel replicas of tuned inference engines across geo-distributed DCs~\cite{AWS, skywalker} and coordinate them through centralized~\cite{central_Spanner}, decentralized~\cite{dynamo_amazon}, or hierarchical~\cite{hierarchical_apollo} control planes. However, as highlighted in~\cite{hierarchical_apollo, casper}, centralized designs are inherently vulnerable to a single point of failure and to bottlenecks under high loads, since each decision passes through the central scheduler, whereas decentralized designs lack a consistent global view and incur significant coordination overhead. Therefore, we adopt a \emph{hierarchical} design, as in several other studies~\cite{casper, carbonAwareLoadBal_1}, in which a \textit{global provisioner} maintains a consistent view of DC states, and \textit{regional schedulers} either place requests on a local DC or forward them to a peer scheduler for inference serving.

\section{Geo-Distributed LLM Serving Model} 
This section formalizes the {\em geo-distributed LLM serving problem}, introduces our hardware-calibrated latency and power models, and presents our simulator, which builds on these models and emulates inference features such as iterative continuous batching.

\subsection{Problem Formulation} \label{sec:problemFormulation}
We model the geo-distributed LLM serving setup with a set of regional schedulers $\mathcal{S}$ and a set of DCs $\mathcal{D}$, where each scheduler $s \in \mathcal{S}$ manages a subset of $\mathcal{D}$, and each DC $d \in \mathcal{D}$ comprises identical nodes, each with a host and multiple GPUs configured as inference engines. A request $r = (t_r^{\mathrm{arr}}, p_r, \mathrm{SLO}_r)$ arrives at its nearest scheduler $s$ at time $t_r^{\mathrm{arr}}$ with a prompt $p_r$ and per-token $\mathrm{SLO}_r$ defined by the service provider for the serving model. The scheduler either dispatches $r$ to a local DC $d$ or forwards it to another scheduler $s'$, which dispatches it to a remote DC $d'$. At the target DC, an \textit{inter-node} scheduler assigns $r$ to a node, whose host selects an inference engine through an intra-node policy, and $r$ waits in that engine's queue. Once admitted for inference, $r$ undergoes a prefill and a series of decode steps until the end-of-sequence (EOS) token, within a batch of other ongoing requests. Despite identical data-parallel inference engines, heterogeneous prompt lengths and dynamic arrivals inherently lead to a load imbalance among them~\cite{Llumnix}. 

Given a request stream $\mathcal{R}$, regional schedulers dispatch requests to DCs to meet their per-token SLOs while minimizing operational carbon emissions. The total emissions across DCs are \scalebox{0.9}{$\mathrm{CO}_2 = \sum_{d \in \mathcal{D}} E_d \cdot \mathrm{CI}_d$}, where \scalebox{0.9}{$\mathrm{CI}_d$} is DC $d$'s grid CI during the serving window. The energy consumption of a DC accumulates over its nodes, each comprising a host and inference engines, as \scalebox{0.9}{$E_d = \mathrm{PUE}_d \cdot \sum_{t} P_d(t) \cdot \delta t$}, where $t$ indexes simulation steps of length $\delta t$. The DC power is \scalebox{0.9}{$P_d(t) = \sum_{n \in d} \big( P^{\mathrm{host}} + \sum_{e \in n} P^{\mathrm{eng}}_e(t) \big)$}, where \scalebox{0.9}{$P^{\mathrm{host}}$} and \scalebox{0.9}{$P^{\mathrm{eng}}_e(t)$} are the host and engine $e$ powers of node $n$ in DC $d$, respectively. Since all compute resources within a DC share the same power source and facility infrastructure, we model each DC with a single PUE and CI~\cite{casper, gcos}. We detail the per-inference step latency and power model (i.e., $P^{\mathrm{host}}$ and $P^{\mathrm{eng}}$) under varying loads in Section~\ref{sec:latencyModeling} and the inference engine's processing behavior in Section~\ref{sec:simulatorDesign}. 

Our hierarchical design decouples scheduling decisions from per-DC state collection via a central \emph{provisioner}, which polls DCs every $\Delta_p$ seconds for i) active batch size $\mathrm{B}_d$, ii) engine queue depth in tokens $\mathrm{Q}_d$, iii) KV-cache occupancy $\mathrm{KV}_d$, and iv) prefill duty $\mathrm{duty}_d$ (i.e., the recent fraction of time spent in prefill), each averaged over the DC's engines, together with the DC's grid CI, $\mathrm{CI}_d$. It broadcasts these as a global snapshot to all schedulers, along with the inter-region latency matrix $\mathrm{RTT}(s, s')$. We intentionally use per-DC averages since statistics at per-node or per-engine granularity would enlarge the polling payload and the schedulers' decision cost.

\subsection{Inference Latency and Power Modeling} \label{sec:latencyModeling} 
LLM inference proceeds in two phases: \emph{prefill} processes all input tokens in one forward pass, and \emph{decode} generates one token per step autoregressively. Both phases pass requests through a stack of mostly identical transformer blocks, whose kernel dimensions are determined by the model architecture, parallelism strategy, and per-step workload~\cite{parallelizationPaper}. Even in mixture-of-experts (MoE) models, which activate experts unevenly, kernels with fixed dimensions (e.g., GEMMs) repeat across blocks and dominate per-step latency~\cite{moe_cost_break}. As inference phases execute the same-dimension blocks sequentially (e.g., 32 times for Llama-3.1-8B), we model prefill and decode step latencies analytically and compose these models with engine-level mechanisms, such as memory constraints and continuous batching, to simulate end-to-end inference.

\begin{figure*}[t]
    \centering
    \includegraphics[width=0.95\linewidth]{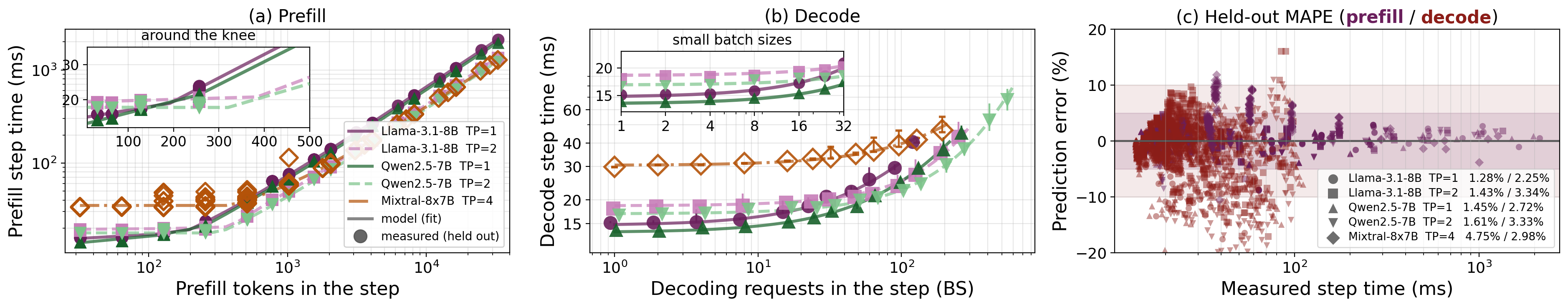}
    \vspace{-2mm}
    \caption{Per-engine latency model across models and TP degrees: \textbf{(a, b)} prefill and decode step time across prefill lengths and batch sizes, respectively; \textbf{(c)} prediction error against measured step time, with per-setup MAPE.}
    \vspace{-5mm}
    \label{fig:inf_latency_val}
\end{figure*}

\noindent {\bf Prefill Latency.} Figure~\ref{fig:inf_latency_val}-a plots prefill time across input tokens for Llama-3.1-8B~\cite{llama3} and Qwen2.5-7B~\cite{qwen} at two tensor-parallel degrees (TP=1 and TP=2), and for Mixtral-8x7B~\cite{mixtral} at TP=4. Prefill latency stays nearly flat up to a few hundred tokens (e.g., 400 for Mixtral-8x7B at TP=4), where GPU compute is not yet saturated, and then grows approximately linearly with the total prefill length. Therefore, for a given model, parallelism strategy, and hardware, prefill latency can be expressed as a function of the total prefill length. We model it as \scalebox{0.9}{$t_{\mathrm{pre}} = t_{\mathrm{tok}}(N) + b_0 + b_1 B + b_2 \textstyle\sum_i n_i^2$}, where $b_0$ and $b_1B$ are the fixed per-step cost and per-request overhead for $B$ requests, and $b_2\sum_i n_i^2$ is the quadratic attention cost across prompt lengths $n_i$. The token term $t_{\mathrm{tok}}(N)$, which captures the cost of processing a total of $N$ tokens, is piecewise linear in $N$, with its slope increasing after the saturation point $k$. We calibrate these coefficients and saturation points separately for each inference setup, using vLLM v0.6.3 measurements in FP16 on A100 GPUs with prompts from diverse datasets~\cite{humanEval, mbpp, govReport} spanning 32 to 32K tokens. The model predicts prefill step time within 1.7\% MAPE for Llama-3.1-8B and Qwen2.5-7B across both TP degrees, and within 4.8\% for Mixtral-8x7B at TP=4 (Figure~\ref{fig:inf_latency_val}-c), whose prompt-dependent expert activations vary the execution time for the same token count.

\noindent {\bf Decode Latency.} Unlike prefill, decode steps generate one token per request by reading model weights and the KV cache, making decode memory-bound~\cite{splitWise}. Batching amortizes each step's weight read across requests, but GPU memory bounds the batch size as KV caches grow. As Figure~\ref{fig:inf_latency_val}-b shows, decode step time grows sublinearly with batch size $B$, as larger kernels improve compute utilization~\cite{parallelizationPaper}, but rises sharply after saturation. It further increases with the cached context because each step reads the KV cache for all batched requests. We capture both trends with a low-order polynomial that is quadratic in $B$, \scalebox{0.9}{$t_{\mathrm{dec}} = a_0 + a_1 B + a_2 B^2$}, where each coefficient varies linearly with the mean cached length per request $\ell = K/B$ as \scalebox{0.9}{$a_j = \alpha_j + \beta_j \ell$} and $K$ is the total number of cached tokens. We calibrate the coefficients using the same inference measurements as in prefill, spanning batch sizes and KV-cache occupancies close to each setup's memory limit. Our model predicts decode step time within 3.4\% MAPE across all five setups (Figure~\ref{fig:inf_latency_val}-c), but since decode steps typically take tens of milliseconds, transient thermal and frequency effects become more visible and limit this accuracy.

\begin{figure}[h]
    \centering
    \vspace{-1mm}
    \includegraphics[width=.92\linewidth]{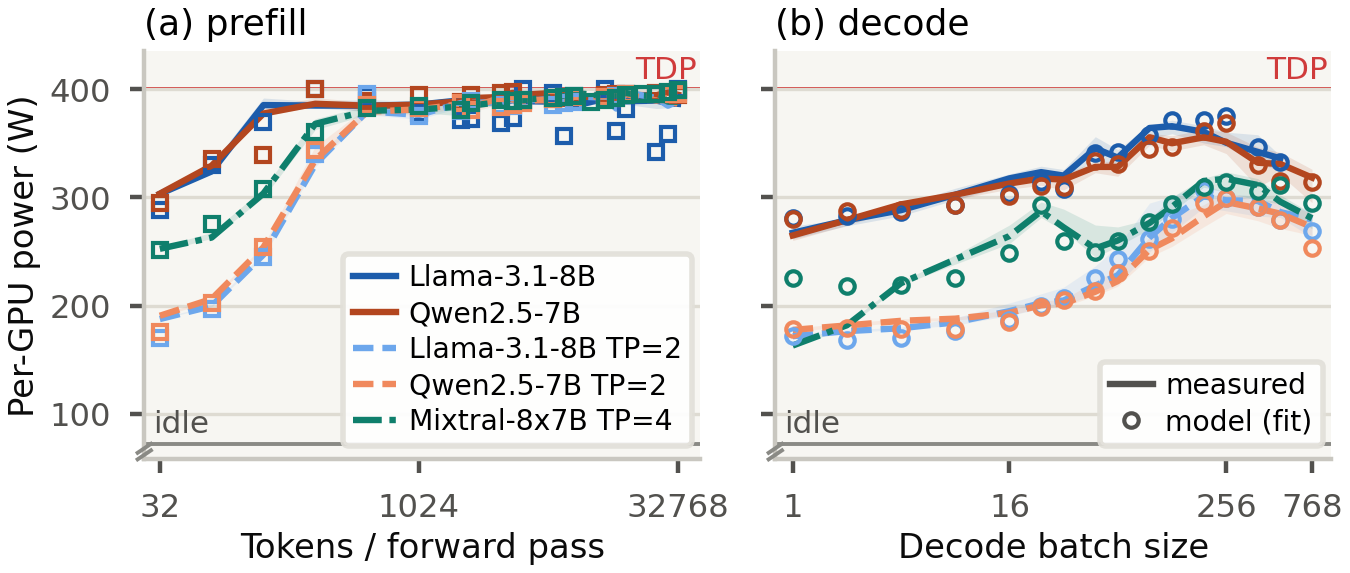}
    \vspace{-2mm}
    \caption{Per-GPU power during \textbf{(a)} prefill and \textbf{(b)} decode across models and TP degrees.}
    \vspace{-1mm}
    \label{fig:power_model}
\end{figure}

\noindent {\bf Power Model.} In addition to the latencies, capturing the dynamic power dissipation during inference is necessary for carbon accounting. Figure~\ref{fig:power_model} shows the measured average per-GPU power across the prefill and decode steps for the same setups. Prefill power starts below the TDP (400\,W for A100~\cite{a100}) for the shortest prompts but quickly reaches it once the tokens saturate compute resources, consistent with Figure~\ref{fig:inf_latency_val}-a. Decode power, instead, grows sublinearly with batch size and reaches about 95\% of TDP for larger batches, with a slight drop at the largest batch sizes. With higher degrees of parallelism, e.g., TP=2 for Llama, the transformer kernels are sharded across GPUs, so per-GPU power starts at a lower value and reaches the TDP at larger input workloads.

We model the energy of each inference step as the average power times the step latency. A prefill step costs \scalebox{0.9}{$E_{\mathrm{pre}} = e^{\mathrm{pre}}_0 + e_p N + e_a \sum_i n_i^2$}, where $e^{\mathrm{pre}}_0$ is the prefill fixed step energy, $e_p N$ is the compute energy of $N$ prefill tokens, and $e_a \sum_i n_i^2$ is the marginal attention energy capturing the effect of the prompt lengths $n_i$. Similarly, a decode step costs \scalebox{0.9}{$E_{\mathrm{dec}} = e^{\mathrm{dec}}_0 + e_d B + e_k K$}, where $e^{\mathrm{dec}}_0$ is the fixed decode step energy,  $e_d B$ is the compute energy of $B$ batched requests, and $e_k K$ captures the energy of reading the $K$ cached tokens. The power model, calibrated per inference setup on the campaigns shown in Figure~\ref{fig:power_model}, predicts held-out step energy within 5.6\% MAPE across all setups. Additionally, the simulator accounts for two standing costs: i) when a model is loaded and ready to serve but no request is in flight (i.e., the \textit{idle} state), an engine draws about 72\,W per A100 GPU, and ii) each node incurs a nearly constant host power $P^{\mathrm{host}}$, which we set to 300\,W, as the CPU package alone draws about 146\,W under engine load on our testbed, and the remainder covers host memory, networking, and peripherals.

\subsection{Geo-distributed LLM Inference Simulator} \label{sec:simulatorDesign}
\begin{figure}[h]
    \centering
    \vspace{-1mm}
    \includegraphics[width=0.91\linewidth]{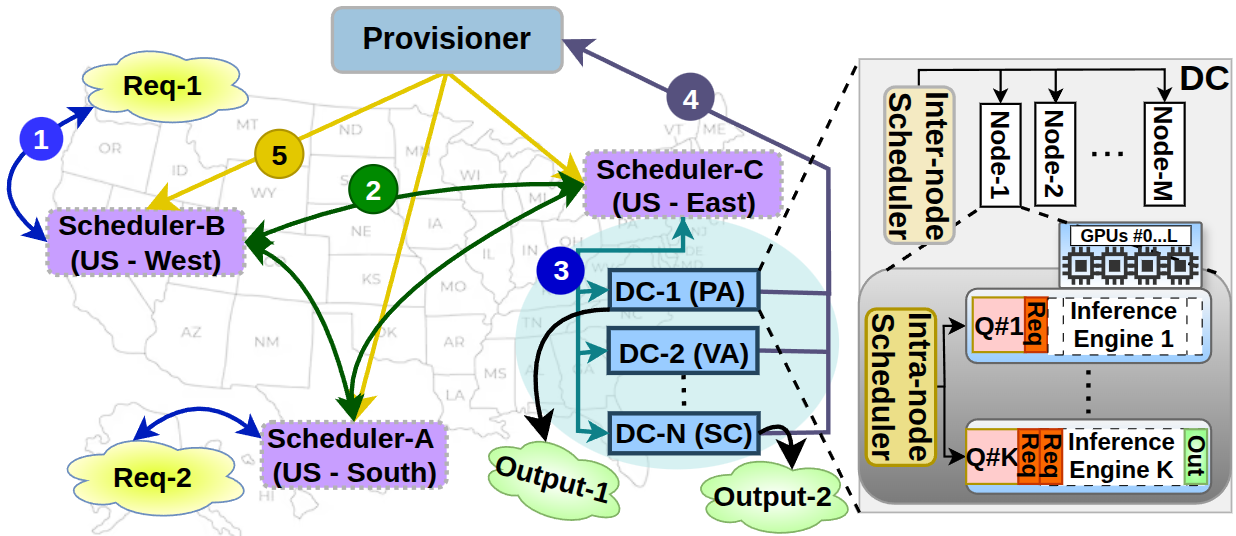}
    \vspace{-2mm}
    \caption{Overview of the simulator architecture on an example topology.} 
    \label{fig:simArchitecture}
    \vspace{-2mm}
\end{figure}

Our simulator, depicted in Figure~\ref{fig:simArchitecture} on an illustrative deployment, consists of four major components communicating over gRPC: (i) a load generator, (ii) regional schedulers, (iii) DCs, and (iv) a provisioner. The load generator, represented in the figure by requests (e.g., Req-1 and Req-2), emits prompts from a dataset and dispatches them to the schedulers~\circled{1} at a configurable Poisson rate $\lambda$, emulating real-world demand with varying prompt lengths and stochastic arrivals.

A regional scheduler (e.g., US-South) schedules each request on a local or remote DC using a policy described in Section~\ref{sec:schedulers}. When the policy selects a remote DC, the scheduler forwards the request to the target region's scheduler~\circled{2}, incurring an inter-region transfer delay, and the remote scheduler admits the request into its scheduling window before placing it on the target DC~\circled{3}. Scheduling policies can optimize a single objective or several at once, such as carbon and latency. Any such policy needs objective-specific state from the DCs. For the carbon--latency trade-off, we design the simulator such that each node continuously maintains per-engine counters for active batch size, queue depth, KV-cache occupancy, and prefill duty. In every polling period $\Delta_p$, the provisioner pulls these counters averaged over each DC's engines, along with the grid CI of the DC~\circled{4}, and broadcasts them to all schedulers with the inter-region latency matrix~\circled{5}. Schedulers consume these to optimize latency and carbon objectives together through distinct policies. We analyze the impact of the polling period on scheduling quality in Section~\ref{sec:results}.

At the target DC, the inter-node scheduler assigns the request to a node, and the intra-node scheduler dispatches it to an inference engine, where it joins the engine's queue (e.g., Q\#1 in Figure~\ref{fig:simArchitecture}). Each engine emulates a continuous-batching loop in wall-clock time, with a configurable maximum batch size similar to vLLM's {\small\texttt{max\_num\_seqs}}~\cite{vllm_engine_args}, and a model-specific constant KV cache size per token (e.g., 128\,KiB for Llama-3.1-8B), determining how many tokens fit into the engine memory left after the model weights. At each inference step, the engine first retires finished requests and then admits queued ones as long as the batch size with the new request stays within its batch limit and the new request's KV cache fits into its remaining memory. Hence, the \textit{queueing delay} depends on the engine's occupancy rather than being a fixed constant. 

Admitted requests can be processed in several prefill steps, depending on the input sequence length (ISL). For example, a prompt whose ISL exceeds the per-step token budget, similar to vLLM's {\small\texttt{max\_num\_batched\_tokens}}~\cite{vllm_engine_args}, is split across consecutive steps, each lasting $t_{\mathrm{pre}}$ and stalling the ongoing decode batch. The updated batch then advances by one decode step lasting $t_{\mathrm{dec}}$, recomputed for the current batch and KV cache size, so that admissions and completions change the inference step time at every iteration. Engines sleep for each predicted step duration, so latencies accumulate in wall-clock time. A request's TTFT thus includes the scheduling delay, any inter-region transfer, queueing time, and prefill duration, whereas its TPOT is the average of the decode step latencies, including those stalled by other requests' prefills. Alongside latency, engines accumulate energy throughout inference, charging $E_{\mathrm{pre}}$ per prefill step, $E_{\mathrm{dec}}$ per decode step, and idle energy when there are no requests in the inference pipeline. Together, the simulator reproduces the latency and energy dynamics of modern inference engines, such as vLLM, which we validate within a node deployment in Section~\ref{sec:sim_validation}.

\section{Carbon-Aware Scheduler Designs} \label{sec:schedulers} 
Building on the simulator's calibrated inference latency and energy models, we investigate the carbon--SLO trade-off through a series of scheduling policies of increasing sophistication, from per-request heuristics to batch-level optimization: (i) a per-request greedy heuristic, (ii) a standard batch-based integer linear program (ILP), (iii) a load-coupled iterative ILP, and (iv) a facility-location mixed-integer linear program (MILP) that prices the energy of batched requests collectively. All four schedulers share a common latency estimator to predict TTFT and TPOT for each request-DC pair $(r, d)$ using the analytical step models given in Section~\ref{sec:latencyModeling}. The estimated TTFT is \scalebox{0.9}{$\mathrm{TTFT}_{r,d} = \mathrm{RTT}(s, s') + W_d + t_{\mathrm{pre}}(\mathrm{ISL}_r + \mathrm{Q}_d)$}, where $\mathrm{RTT}(s, s')$ is the inter-region transfer delay (zero for a local DC placement), $W_d$ is $r$'s wait time for a batch slot at an engine in $d$, and $t_{\mathrm{pre}}(\mathrm{ISL}_r + \mathrm{Q}_d)$ is the prefill time of $r$ together with $\mathrm{Q}_d$ previously queued prompt tokens. The estimated TPOT is \scalebox{0.9}{$\mathrm{TPOT}_{r,d} = \big[t_{\mathrm{dec}}(\mathrm{B}_d + 1,\, \mathrm{KV}_d) + T^{\mathrm{pre}}_{d}\big]/(1 - \mathrm{duty}_d)$}, where $\mathrm{B}_d$ and $\mathrm{KV}_d$ are the average batch size and KV occupancy per engine in $d$, with $+1$ accounting for $r$ joining the batch, the factor \scalebox{0.9}{$1/(1 - \mathrm{duty}_d)$} for prefills already running at $d$, and $T^{\mathrm{pre}}_{d}$ for those still queued, amortized over $r$'s output length. Both latency estimates rely on inputs available at scheduling time: $\mathrm{ISL}_r$ from tokenizing $r$'s prompt, and $\mathrm{B}_d$, $\mathrm{Q}_d$, $\mathrm{KV}_d$, $\mathrm{duty}_d$, and $\mathrm{RTT}(s, s')$ from the provisioner's latest snapshot.

The TPOT (only in $T^{\mathrm{pre}}_d$) and carbon estimates require $r$'s output sequence length (OSL), which is unknown until inference ends. For example, a request with an OSL of 100 occupies an engine roughly $10\times$ longer than one with an OSL of 10 and consumes proportionally more decode energy, so an approximate OSL matters for placement quality. We evaluate the schedulers with both \textit{oracle} and \textit{predicted} OSLs, where the latter is the median quantile estimated by a lightweight two-layer MLP (detailed in Section~\ref{sec:results}) from the request's ISL, serving model, and application type. With these estimates, the scheduler compares the predicted TTFT and TPOT with the serving model's budgets to identify SLO-feasible DCs. The following four policies then trade carbon against latency among the feasible DCs, differing in their cost models, how they project load onto DCs, and the granularity at which they schedule requests.

\subsection{Carbon-Aware Greedy Scheduling} 
The greedy policy makes an independent scheduling decision for each request using a \textit{scoring} function. For an incoming request $r$, the scheduler first filters out DCs that violate the per-token SLOs via the analytical TTFT and TPOT estimates, and then scores the remaining SLO-feasible DCs, $\mathcal{F}_r$, with:

{\small\setlength{\abovedisplayskip}{-1pt}\setlength{\belowdisplayskip}{7pt}
\begin{equation} 
\mathrm{score}_{r,d} = \alpha \, \frac{\mathrm{Lat}_{r,d}}{\max_{d' \in \mathcal{F}_r} \mathrm{Lat}_{r,d'}}
+ (1-\alpha) \, \frac{\mathrm{CI}_d}{\max_{d' \in \mathcal{F}_r} \mathrm{CI}_{d'}}
\label{eq:greedy_score}
\end{equation}}%
where $\mathrm{Lat}_{r,d} = \mathrm{TTFT}_{r,d} + \mathrm{OSL}_r \cdot \mathrm{TPOT}_{r,d}$ is the estimated end-to-end latency of $r$ at $d$, and $r$ is dispatched to the lowest-scoring DC. The trade-off parameter $\alpha \in [0,1]$ weighs latency against carbon, with $\alpha{=}1$ prioritizing latency and $\alpha{=}0$ prioritizing carbon. The key weakness of the greedy policy is that it scores each request in isolation and cannot coordinate placements across requests. Thus, within a polling period, it keeps dispatching requests to the same carbon-latency optimal DC until the projected load drives that DC to the SLO boundary. Since the regional schedulers operate concurrently on the same snapshot, their combined dispatches easily push the DC beyond this boundary, violating request SLOs.

\subsection{Carbon-Aware Batch Scheduling via ILP} 
In contrast to greedy per-request scheduling, the batch schedulers \emph{aggregate} requests over a short window $\Delta_w$ or up to a fixed count and jointly assign them by solving an ILP, which limits the load placed on any DC and removes herding among a scheduler's own requests. Let $\mathcal{R}$ denote the requests accumulated in $\Delta_w$, and let $x_{r,d} \in \{0,1\}$ indicate whether $r$ is assigned to $d \in \mathcal{F}_r$, where $\mathcal{F}_r$ is the SLO-feasible DC set for $r$. The assignment is obtained by solving:

{\small\allowdisplaybreaks\setlength{\abovedisplayskip}{-2pt}\setlength{\belowdisplayskip}{8pt}
\begin{subequations}\label{eq:std_ilp}
\begin{align}
\min_{\mathbf{x}} \ & \sum_{r \in \mathcal{R}} \sum_{d \in \mathcal{F}_r} c_{r,d}\, x_{r,d} \label{eq:std_obj}\\
\text{s.t.}\ & \sum_{d \in \mathcal{F}_r} x_{r,d} = 1, \quad \forall\, r \in \mathcal{R} \label{eq:std_assign}\\
& \sum_{r \in \mathcal{R}} x_{r,d} \le H_d, \quad \forall\, d \in \mathcal{D} \label{eq:std_cap}
\end{align}
\end{subequations}}%
where \scalebox{0.9}{$c_{r,d} = \mathrm{PUE}_d \cdot E_{r,d} \cdot \mathrm{CI}_d$} is the carbon cost of serving $r$ at $d$, and $E_{r,d}$ is the energy consumed for serving $r$ at $d$ under the power model of Section~\ref{sec:latencyModeling} (i.e., the increase in step energy from its prefill and its $\mathrm{OSL}_r$ decode steps). \scalebox{0.9}{$H_d = \max(0, \mathrm{cap}_d -$} \scalebox{0.9}{$\mathrm{load}_d)$} represents the remaining slots in $d$'s total batch capacity $\mathrm{cap}_d$ (i.e., the per-engine batch capacity times its engine count), with $\mathrm{load}_d$ corresponding to active and pending requests in the projected view of $d$. Since each $x_{r,d}$ appears in one assignment equality and one capacity inequality, the constraint matrix is totally unimodular, and the LP relaxation yields integral optima. We solve the ILP as a linear program with HiGHS~\cite{highs} within tens of milliseconds.

The \textbf{standard ILP} solves this program once per window, estimating each request's feasibility against the same pre-window state without accounting for the other requests placed in the window. Hence, a DC feasible for each request alone may violate SLOs once they are placed together, as their combined batch and queue push TTFT and TPOT beyond the estimates. To address this load-coupling gap, we formulate the \textbf{iterative ILP}, which feeds each solution back into the view. After each iteration $t$, the assignments $n^{(t)}_d = \sum_r x^{(t)}_{r,d}$ are projected onto each DC's batch size and queue as if already dispatched, and \scalebox{0.9}{$\mathrm{TTFT}^{(t)}_{r,d}$} and \scalebox{0.9}{$\mathrm{TPOT}^{(t)}_{r,d}$} are recomputed with the updated states to refresh \scalebox{0.9}{$\mathcal{F}^{(t)}_r$} and \scalebox{0.9}{$E^{(t)}_{r,d}$}. Equation~\eqref{eq:std_ilp} is then re-solved until two consecutive iterations yield the same assignment or $T_{\max}$ rounds elapse, at the cost of several solves per window.

\subsection{Carbon-Aware Batch Scheduling via MILP}
However, the ILP-based schedulers account for interactions among requests assigned to the same DC only through capacity and latency constraints while treating their carbon costs independently, as in prior per-request formulations~\cite{greenllm,moerRouting}. Consequently, they estimate the carbon cost of serving multiple requests as the sum of their individual costs, overlooking the shared energy consumption of batched inference. In practice, requests within the same batch share an inference engine's energy, including a fixed per-step energy cost that does not scale with batch size. For example, when two requests are assigned to an idle engine, an additive model charges the engine's activation cost twice, even though both requests can share a single batch and incur that cost only once.

To address this, we formulate a \textbf{facility MILP} that separates the fixed carbon a DC pays once if it serves any request in the window from the per-request cost $c_{r,d}$. It adds $\sum_d f_d\, y_d$ to the objective in Equation~\eqref{eq:std_obj} and replaces the capacity constraint in Equation~\eqref{eq:std_cap} with $\sum_r x_{r,d} \le H_d\, y_d$, where $y_d \in \{0,1\}$ indicates whether $d$ receives any requests from the window, and $f_d$ is the carbon of waking $d$'s engines. With this capacitated facility MILP, the wake-up cost is charged once per DC, regardless of how many requests are placed there. Although harder to solve than the ILP, the MILP adds only $|\mathcal{D}|$ binaries (i.e., one $y_d$ per DC) compared with the $|\mathcal{R}| \cdot |\mathcal{D}|$ assignment variables, so HiGHS solves it within the 0.25\,s time limit. We evaluate the carbon--SLO trade-offs of all four schedulers, together with their scheduling overheads, in Section~\ref{sec:results}.

\section{Experimental Evaluation}
\subsection{Experiment Setup} \label{sec:expSetup} 

\noindent\textbf{Geo-distributed setup.} We evaluate our schedulers across four regions (EU-North, US-East, Asia, and South America), each with four DCs (e.g., Netherlands, Belgium, Lithuania, and Latvia for EU-North) whose CI is drawn from Electricity Maps~\cite{elecMaps}. Each DC comprises four nodes of 8$\times$A100-40GB GPUs~\cite{a100}, and its hourly CI (i.e., the finest available granularity) is held constant within a run. The inter-region transfer delay is the measured one-year p50 RTT between the corresponding AWS regions~\cite{cloudping}. We use a uniform PUE of 1.1~\cite{google_pue} by default and assess real-world sensitivity with per-DC PUE values from public disclosures~\cite{realPUE}. As three DCs lack public PUE data, the PUE, CI, and topology studies replace them and their CIs with the nearest-CI sites in each region. Within DCs, both inter-node and intra-node schedulers use round-robin so that the measured carbon savings stem solely from the regional schedulers' decisions.

\begin{table}[t]
\centering\footnotesize
\setlength{\tabcolsep}{3pt}
\caption{Topology, serving, and scheduler parameters.}
\label{tab:params}
\begin{tabularx}{\columnwidth}{@{}l>{\raggedright\arraybackslash}X@{}}
\toprule
\textbf{Parameter} & \textbf{Value} \\
\midrule
\multicolumn{2}{@{}l}{\textit{\textbf{Geo-distributed topology}}} \\
Schedulers / DCs per scheduler & 4 / 4 \\
Nodes per DC / GPUs per node & 4 / 8$\times$A100-40GB \\
Inter-region RTT & 85--280\,ms, AWS p50~\cite{cloudping} \\
$\mathrm{PUE}$ across DCs & 1.1 uniform (real per DC in sensitivity) \\
\midrule
\multicolumn{2}{@{}l}{\textit{\textbf{Serving configuration}}} \\
Llama-3.1-8B (Dense)~\cite{llama3} & 32 layers, 8 KV heads, TP=1 (TP=2 in Sec.~\ref{sec:discussion}) \\
Qwen2.5-7B (Dense)~\cite{qwen} & 28 layers, 4 KV heads, TP=1 \\
Mixtral-8x7B (MoE)~\cite{mixtral} & 32 layers, 8 KV heads, top-2 of 8 experts (13B active), TP=4 \\
Arrival rate $\lambda$ & 1200--4000 req/s (Mixtral 800--2000; long/mixed 50/200) \\
TTFT / TPOT budget & 3000 / 100\,ms (Mixtral TPOT 200\,ms; long/mixed TTFT 4500\,ms) \\
\midrule
\multicolumn{2}{@{}l}{\textit{\textbf{Scheduler configuration}}} \\
$\Delta_p$, poll mode & 1\,s (swept 0.5--10\,s, sync and async) \\
$\Delta_w$, request limit & 100\,ms, 300 req. \\
Iter. ILP $T_{\max}$ / MILP time limit & 3 / 0.25\,s \\
Per-engine batch capacity & 40 (TP=1), 80 (TP=2), 400 (Mixtral) \\
Greedy $\alpha$ & 0, 0.5, 0.9 \\
$\mathrm{OSL}_r$ source & both oracle and predicted \\
\bottomrule
\end{tabularx}
\vspace{-5mm}
\end{table}

\noindent {\bf Simulator configuration.} We evaluate the proposed schedulers on geo-distributed deployments of Llama-3.1-8B at TP=1 and Mixtral-8x7B at TP=4, for which the MLPerf Inference \textit{Server} scenario~\cite{mlperfInfer} sets (TTFT, TPOT) SLOs of (2\,s, 100\,ms) on CNN-DailyMail~\cite{cnn_daily} and (2\,s, 200\,ms) on its OpenOrca-based workload~\cite{OpenOrca}, respectively. Since MLPerf evaluations typically run on a single node, we relax both TTFT budgets to 3\,s to account for inter-region forwarding, scheduling, and inter-node dispatch delays, while keeping the TPOT budgets unchanged. We further evaluate Qwen2.5-7B~\cite{qwen} under the Llama-3.1-8B trace and SLOs to test whether our findings generalize beyond the MLPerf models. The load generator emits requests with Poisson arrivals at the rates in Table~\ref{tab:params} and dispatches them round-robin to the four schedulers for 200\,s per policy-rate pair on a single 64-core server, allowing engines to reach a stable load within the first 15\,s, leaving over 90\% of each run at steady-state occupancy. We define the load regime by the ratio of the mean per-engine batch size under round-robin to the batch capacity: i) below 15\% as \textit{low}, ii) 15--40\% as \textit{moderate}, and iii) above 40\% as \textit{high} load. Accordingly, we use 1200, 2400, and 4000 req/s as the low, moderate, and high loads for the 7B and 8B models, and 800, 1200, and 2000 req/s for Mixtral-8x7B.

\noindent \textbf{Scheduler configuration.} We sweep the greedy policy over the $\alpha$ values in Table~\ref{tab:params}, from carbon-only ($\alpha{=}0$) to latency-leaning ($\alpha{=}0.9$). Both ILP- and MILP-based policies batch requests over a 100\,ms window that closes early at 300 requests, with at most $T_{\max}{=}3$ iterations for the iterative ILP and a 0.25\,s solver limit for the MILP. When a solver fails or cannot place a request within capacity, the scheduler falls back to the lowest-carbon SLO-feasible DC with headroom (i.e., greedy with $\alpha{=}0$), and to round-robin (RR) if none exists. We use oracle OSLs in all experiments except the OSL sensitivity study. As carbon-blind baselines, we test RR and least-loaded (LL), which schedule each request individually. We further compare our schedulers against Green-LLM~\cite{greenllm}, with its hourly per-DC token budget mapped to the engine batch limit, and MOER-based routing~\cite{moerRouting}, porting only their scheduling decisions into our system. MOER-based routing sends each request to the lowest-CI SLO-feasible DC, and thus coincides with our greedy policy at $\alpha{=}0$.

\noindent\textbf{Evaluation metrics.} Our primary metrics are total operational carbon in gCO$_2$, including idle and host energy, and per-token SLO violation, defined as the fraction of requests that miss either their TTFT or TPOT budget. Carbon savings are measured against RR at the same load. Since SLO violation rates hide tail latency, we additionally analyze the p50, p90, and p99 TTFT across policies, as well as the per-request scheduling overhead.

\subsection{Simulator End-to-End Validation} \label{sec:sim_validation}
Beyond the per-step analytical models in Section~\ref{sec:latencyModeling}, we validate the simulator against real end-to-end inference serving. We deploy three data-parallel Llama-3.1-8B vLLM engines (the same version as in calibration) at TP=1 on A100-40GB GPUs and dispatch CNN-DailyMail prompts to them with the RR policy at Poisson arrival rates from 2 to 50\,req/s. The rates span 7\% to 181\% of the engines' measured capacity (27.58\,req/s under saturation), covering both the queue-free and queueing regimes. We then replay each run through the simulator with all decisions fixed, sending every request to the same engine at its logged arrival time with its logged ISL and actual OSL, so that the comparison isolates the engine model.

Across the 19 rates and over 40K requests, the simulator captures the TPOT latencies that vary from 15.5\,ms to 54\,ms as prefills increasingly stall decode steps, with an average error of less than 5\%. It further reproduces TTFT latencies, ranging from 71\,ms to over 20\,s as queueing builds up, with an overall average error of 9.1\%, which varies from 3.5\% below saturation and 5.2\% under overload to 28.5\% around saturation, where it conservatively overestimates TTFT. Running a single engine instead of three of them at the same per-engine load changes TPOT by only 3.3\%, confirming that co-located engines do not interfere with each other and can be modeled independently. In addition, the simulated GPU energy stays within 2.5\% of the measurement at all arrival rates, showing that the simulator closely captures the end-to-end engine behavior from queue-free operation to overload.

\subsection{Carbon--SLO Characterization} \label{sec:results}
\begin{figure}[h]
    \centering
    \subfloat[Carbon saving relative to RR and SLO violation for each policy.\label{fig:slo_vs_carbon}]{%
        \includegraphics[width=\linewidth]{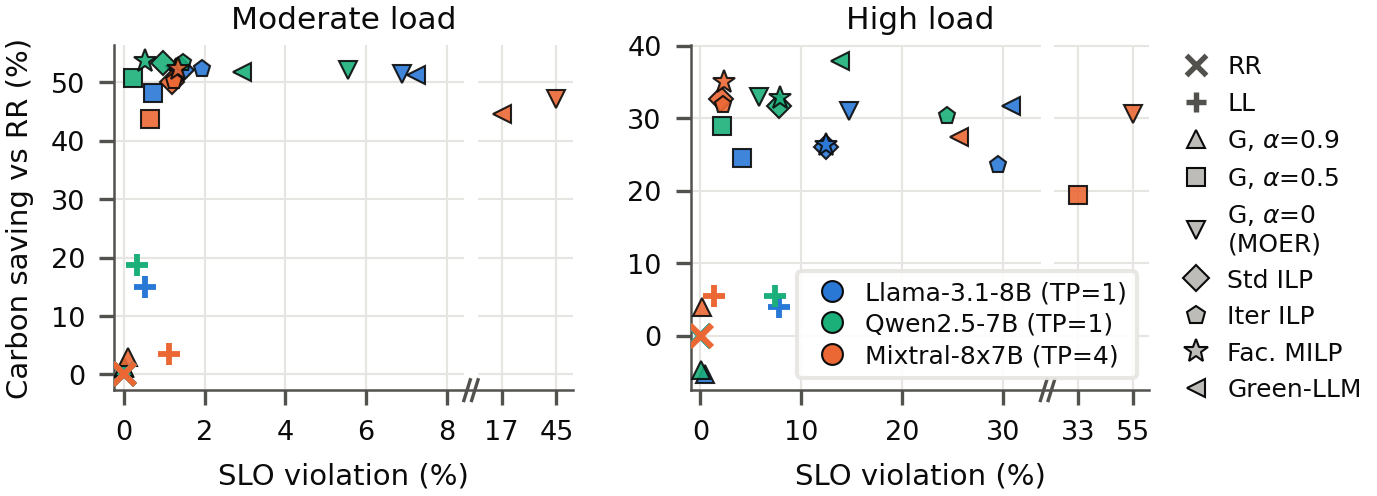}
    } \\[8pt]
    \subfloat[TTFT tail behavior per policy at \textit{high} load, from p50 to p99 with ticks at p50/p90. Labels show carbon savings relative to RR.\label{fig:tail_vs_carbon}]
    {%
        \includegraphics[width=\linewidth]{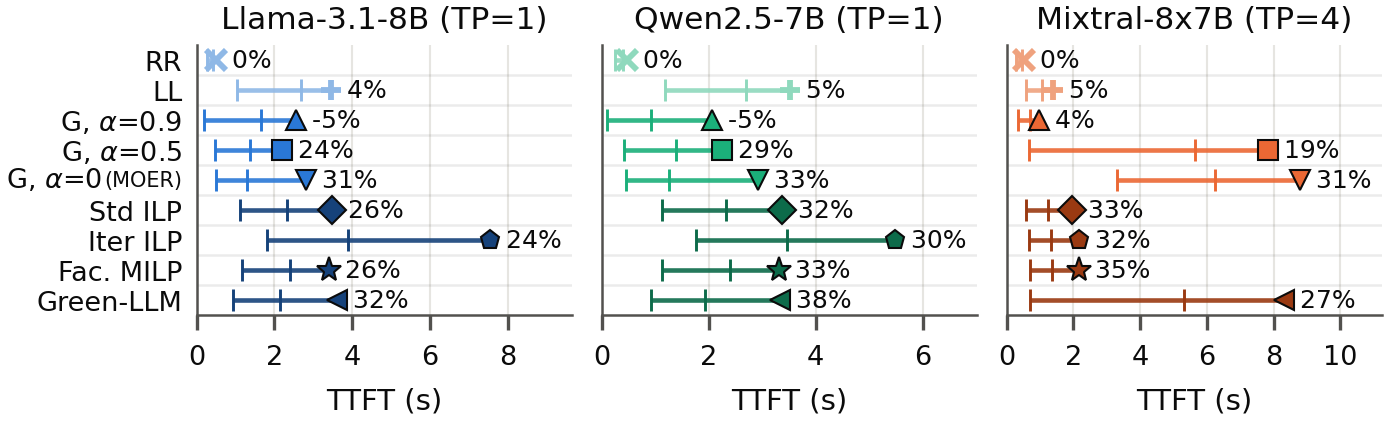}}
    \caption{Carbon--SLO trade-off and TTFT tail behavior across scheduling policies, loads, and models. G denotes greedy policies, with $\alpha{=}0$ corresponding to the MOER~\cite{moerRouting} policy.}
    \vspace{-4mm}
    \label{fig:policy_comparison}
\end{figure}

Figure~\ref{fig:policy_comparison} presents the carbon--SLO trade-off across scheduling policies at moderate and high loads, including SLO violations against normalized carbon savings relative to RR and TTFT tail behavior at high load across all models. We omit the low-load case from the figure, where all carbon-aware schedulers except greedy at $\alpha{=}0$ and Green-LLM~\cite{greenllm} remain within 1.3\% SLO violation, and the facility MILP achieves the highest carbon savings across all models (53.8--59.6\%).

\noindent\textbf{Carbon--SLO violation trade-offs.} As Figure~\ref{fig:slo_vs_carbon} shows, the greedy policies exhibit $\alpha$-monotonicity, with savings rising and SLO violations growing as $\alpha$ decreases. At moderate system loads, greedy at $\alpha{=}0$ (equivalent to MOER~\cite{moerRouting}) achieves carbon savings similar to those of the ILP- and MILP-based policies but violates the SLO for up to 6.9\% of the requests on the dense models and 44.8\% on Mixtral-8x7B. This is because it concentrates the load on the greenest DCs without a capacity limit. Conversely, greedy at $\alpha{=}0.9$ saves at most 3\% carbon with negligible SLO violation. The ILPs, the facility MILP, and greedy at $\alpha{=}0.5$ perform similarly on the dense models, achieving up to 54\% carbon savings within 1.9\% violation, whereas Green-LLM~\cite{greenllm} reaches comparable savings with up to 7.2\% violation, as its batch-agnostic latency estimates never reject a loaded DC. On Mixtral-8x7B, however, the facility MILP stands out, saving up to 8.6 percentage points more than these policies at a 1.3\% SLO violation.

At high system loads, the highest savings shrink to 32--38\% as the green DCs saturate. Greedy at $\alpha{=}0$ saves 7.3 percentage points more than $\alpha{=}0.5$ on average, but incurs 12.1 points more violations, with the gap widening from Qwen2.5-7B to Llama-3.1-8B to Mixtral-8x7B. This ordering follows each model's per-step decode latency (Figure~\ref{fig:inf_latency_val}-b), which leaves progressively less TPOT headroom under the RR policy (from 71\% to 51\% of the budget), and concentrating the load on greener DCs thus results in more SLO violations. Furthermore, greedy at $\alpha{=}0.9$ emits more carbon than RR on the dense models, as its latency term redirects requests from the loaded green DCs to dirtier ones. In contrast, the standard ILP and the facility MILP remain robust owing to their per-window capacity constraints, saving up to 32.8\% on dense models and 35.0\% on Mixtral-8x7B, with SLO violation rates of up to 12.5\% and 2.4\%, respectively. Green-LLM saves about 5 percentage points more on dense models but substantially worsens SLO violations (up to 30.7\%), and on Mixtral-8x7B, it saves 7.6 percentage points less than the facility MILP, with a violation rate of 25.6\%.

\noindent\textbf{Tail latency across policies.} Since the SLO violation rate hides the severity of violations, we break down the TTFT tail behaviors at high system loads, where queueing pressure is most severe, in Figure~\ref{fig:tail_vs_carbon}. On the dense models, the greedy policies achieve the lowest median TTFT (i.e., p50) among the carbon-aware schedulers (around 0.5\,s), as they place each request upon arrival without a batch capacity constraint. The standard ILP and the facility MILP raise the median up to 1.2\,s, mostly because their capacity constraints cluster arrivals at the DCs, doubling the prompts fused per prefill step, on top of the solver overhead. The iterative ILP raises the median to 1.8\,s due to repeated solves, while Green-LLM reaches only 0.9\,s by avoiding clustered arrivals but keeping the green engines at their batch limit, which increases the queueing wait. The p99 TTFT stays within the 3\,s budget under the greedy policies, but reaches up to 3.6\,s with the standard ILP, the facility MILP, and Green-LLM, and further climbs to 7.5\,s with the iterative ILP due to solver overheads. This indicates that greedy at $\alpha{=}0$ violates the SLOs through TPOT rather than TTFT, as it concentrates decode batches on the greenest DCs. 

In contrast, the Mixtral-8x7B setup at TP=4 runs a quarter as many engines per DC as the dense models at TP=1 and serves about three times longer outputs (a mean OSL of 211 compared to 71 tokens for Llama-3.1-8B). Greedy without a batch capacity constraint thus quickly saturates the green DCs, raising the median TTFT to 3.3\,s at $\alpha{=}0$ due to an average queueing time of 2.9\,s, whereas the ILP- and MILP-based schedulers keep the median within 0.7\,s and their queueing time around 0.2\,s. The p99 TTFT reaches up to 8.8\,s with the carbon-aggressive greedy policies (i.e., $\alpha{\le}0.5$) and Green-LLM but remains within 2.2\,s under the ILP- and MILP-based schedulers, showing that the capacity constraints enforced by these policies matter most for the model with the smallest TPOT headroom.

\begin{figure}[h]
    \centering
    \vspace{-1mm}
    \includegraphics[width=0.92\linewidth]{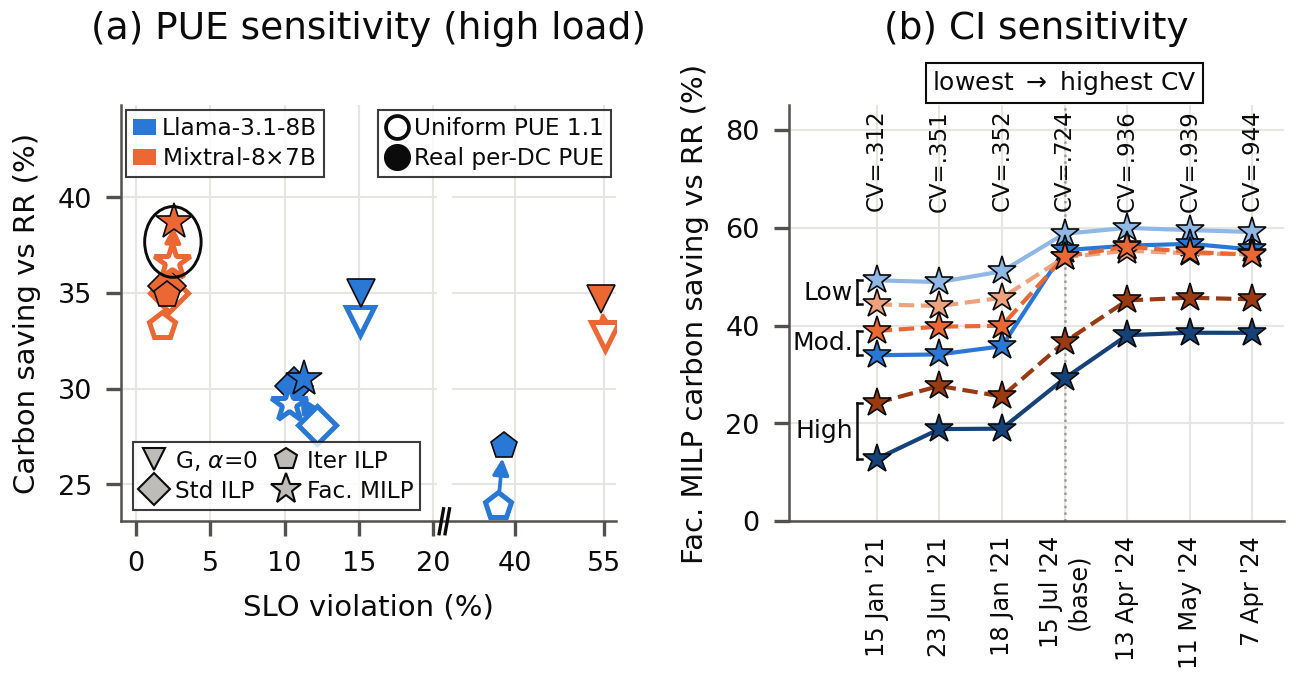}
    \vspace{-3mm}
    \caption{Carbon saving and SLO violation at \textit{high} load under uniform and real per-DC PUE~\cite{realPUE} \textbf{(a)}. \textit{Facility MILP} savings at the base hour and three lowest- and highest-CV hours (CV of CI across DCs) from 2021 to 2024, under varying loads \textbf{(b)}.}
    \vspace{-2mm}
    \label{fig:pue_ci_sensitivity}
\end{figure}

\noindent{\bf PUE and CI Sensitivity.} In practice, PUE varies across DCs and CI varies across regions and over time, making carbon savings sensitive to both. To analyze the sensitivity of our findings to real per-DC PUE, we replace the uniform PUE of 1.1 with operator-reported per-DC values, ranging from 1.08 (e.g., Netherlands) to 1.32 (e.g., Japan), and evaluate the most carbon-aggressive greedy, ILP, and MILP-based policies. As Figure~\ref{fig:pue_ci_sensitivity}-a shows, real per-DC PUEs increase the savings of all carbon-aware schedulers at high load, with the facility MILP improving from 29.2\% to 30.5\% on Llama-3.1-8B and from 36.6\% to 38.7\% on Mixtral-8x7B. In the real PUE data, the dirtiest grids have the highest PUE (e.g., the Japanese DCs combine a CI of 449--502\,gCO$_2$/kWh with a PUE of 1.32), which inflates the carbon of RR more than that of the carbon-aware schedulers. However, their SLO violation remains nearly unchanged, as expected, since PUE mainly rescales each DC's energy use rather than changing where requests are placed. 

Furthermore, Figure~\ref{fig:pue_ci_sensitivity}-b shows how carbon savings respond to the spatiotemporal variation of CI. To quantify this variation, we compute the \textit{coefficient of variation} (CV), i.e., the standard deviation of CI divided by its mean, across DCs at each hour in our CI data. Among the 35K hours from 2021 to 2024, we select the three lowest- and highest-CV hours as well as the base hour and test the \textit{facility MILP} scheduler at those times. From the lowest (CV of 0.31) to the highest-CV hour (CV of 0.94), carbon savings rise from 49/44\% to 59/55\% at low load, from 34/39\% to 56/55\% at moderate load, and from 13/24\% to 39/45\% at high load on Llama-3.1-8B/Mixtral-8x7B models, while the SLO violation remains unchanged. Even though low and moderate system loads leave more room for carbon savings (as more engines in the green DCs remain underutilized), the facility MILP still saves up to 39--45\% carbon across models at high load when the CI spread across DCs is large.

\begin{figure}[h]
    \centering
    \vspace{-1mm}
    \includegraphics[width=0.95\linewidth]{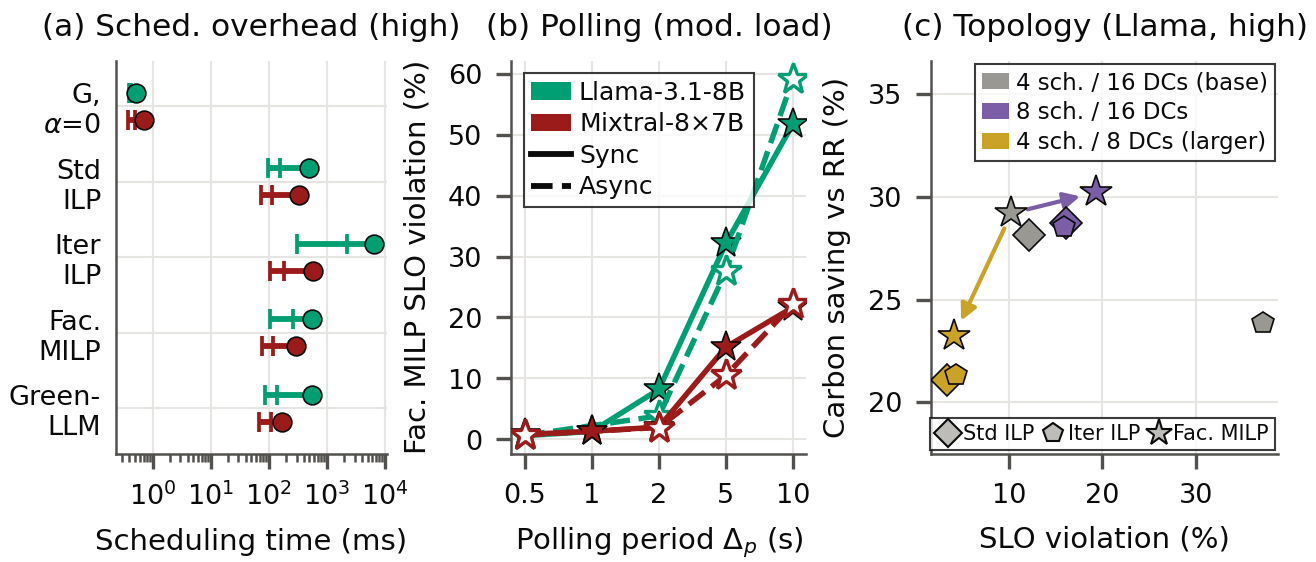}
    \vspace{-2mm}
    \caption{Per-request scheduling time at \textit{high} load, spanning p50 to p99 (\textbf{a}). SLO violation at \textit{moderate} load across polling periods, under \textit{synchronous} and \textit{asynchronous} polling \textbf{(b)}. Carbon saving and SLO violation when serving Llama-3.1-8B at \textit{high} load across various topologies \textbf{(c)}.}
    \vspace{-1mm}
    \label{fig:system_sensitivity}
\end{figure}

\noindent{\bf Scheduler Overhead, Polling Period, and Topology Sensitivity.} As schedulers differ in decision granularity and solver invocations, we compare their per-request scheduling overhead at high load in Figure~\ref{fig:system_sensitivity}-a. The greedy policies decide in under 1\,ms, while the standard ILP, the facility MILP, and Green-LLM show similar overheads with a p99 of up to 0.5\,s. The iterative ILP incurs the highest overhead (a p99 of 6.5\,s on Llama-3.1-8B), as its repeated solves saturate the solver at 1000\,req/s per scheduler. Overall, despite solving a mixed-integer program, the facility MILP offers the best carbon--SLO balance at a comparable overhead. Moreover, these schedulers rely on the DC loads periodically broadcast by the provisioner, so the freshness of this snapshot is critical. To investigate how snapshot freshness affects SLO violation, we vary the polling period under both \textit{synchronous} and \textit{asynchronous} polling in Figure~\ref{fig:system_sensitivity}-b. With a polling period of up to 2\,s, the facility MILP keeps the SLO violation below 10\% for both models (at most 8.2\%), but at 5\,s, the violation rises to 10.5--32\%. Hence, the freshness of load feedback directly bounds the SLO attainment in geo-distributed scheduling. Furthermore, asynchronous polling offers no consistent benefit over synchronous, as staggered polling only spreads the snapshots over time without reducing their age.

Additionally, we study how topology design affects the carbon--SLO trade-off. Beyond the base topology, we design a topology with 8 schedulers, halving the DCs per scheduler, and another with 4 schedulers that halves the total number of DCs while doubling the nodes per DC, so that all topologies have the same number of engines. As Figure~\ref{fig:system_sensitivity}-c shows for the facility MILP on Llama-3.1-8B, doubling the schedulers keeps the carbon savings almost unchanged but nearly doubles the SLO violation (10.3\% to 19.4\%), as more schedulers concurrently concentrate load on the same green DCs from the same snapshot. In contrast, fewer but larger DCs reduce the violation to 4.1\% as DCs absorb more load, but lower the savings to 23.2\% since fewer green DCs remain to schedule requests on.

\begin{figure}[h]
    \centering
    \vspace{-4mm}
    \includegraphics[width=0.92\linewidth]{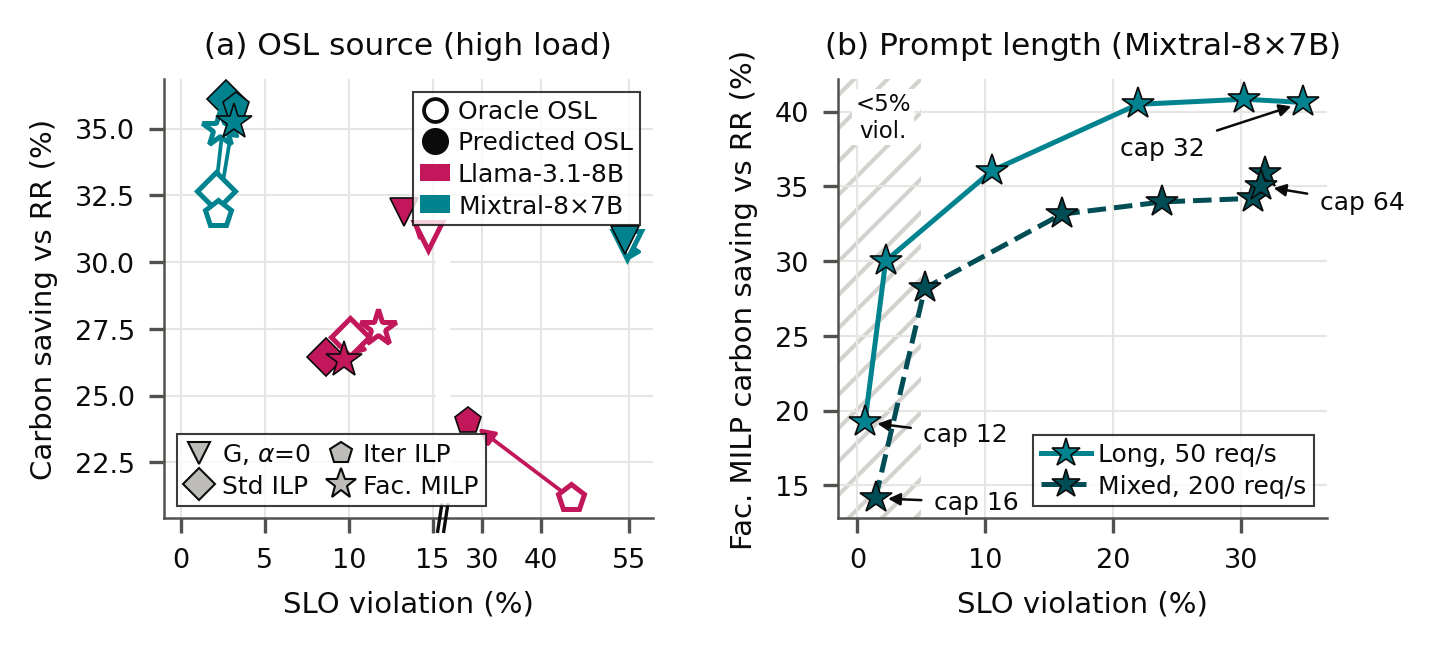}
    \vspace{-3mm}
    \caption{Carbon saving and SLO violation at \textit{high} load with oracle and predicted OSLs \textbf{(a)}. Facility MILP on Mixtral-8x7B with long and mixed prompts as the per-engine batch capacity grows in steps of 4 and 8, respectively \textbf{(b)}.}
    \vspace{-1mm}
    \label{fig:osl_longmixed}
\end{figure}

\noindent{\bf OSL Prediction and Prompt Length Sensitivity.} Since the OSL of a request is unknown at scheduling time, we study replacing the oracle OSL with the OSL estimate of a lightweight predictor, following prior length-aware schedulers~\cite{OSLpred}. We train a two-layer MLP (128 hidden units, about 27K parameters) that takes the log ISL, the serving model, and the dataset (i.e., the application type) as inputs and runs in under 0.1\,ms per request. It jointly classifies each request into one of \textit{six OSL buckets} and estimates its OSL quantiles, and the scheduler uses the median quantile as the OSL. Although only 51\% (Llama-3.1-8B) and 37\% (Mixtral-8x7B) of the estimates fall into the correct bucket (compared to 17\% for a random guess), 99\% and 70\% fall in the correct or an adjacent bucket. As Figure~\ref{fig:osl_longmixed}-a shows, predicted OSLs barely move any scheduler, changing carbon savings by at most 1.3 points on Llama-3.1-8B and 4.1 points on Mixtral-8x7B, and raising SLO violations by at most 1.1 points, except for the iterative ILP on Llama-3.1-8B, whose results vary with its decision latency.

We further evaluate sensitivity to prompt length, since the MLPerf workloads have short-to-moderate ISLs (a mean of 841 tokens for CNN-DailyMail and 195 for OpenOrca). We use a long set of prompts drawn from GovReport~\cite{govReport} and a mixed set of prompts drawn from six datasets~\cite{govReport,cnn_daily,OpenOrca,gsm8k,humanEval,mbpp} with a mean ISL of 10.7K and 2.9K tokens, respectively, on Mixtral-8x7B. Since longer prompts occupy more KV cache per request, the batch capacity that an engine can sustain changes. Thus, we sweep the per-engine batch capacity and relax the TTFT budget to 4.5\,s for both scenarios as long prompts reach a p99 TTFT of 1.7\,s even at low load under the RR policy. We choose the rates (50 and 200 req/s) so that RR fills about 40\% of the per-engine KV cache, the moderate-to-high boundary measured by KV occupancy. As Figure~\ref{fig:osl_longmixed}-b shows, the facility MILP saves up to 30\% carbon at a batch capacity of 16 on long prompts and 28\% at 24 on mixed prompts, while keeping the SLO violation near 5\%, indicating that carbon-aware scheduling generalizes across context lengths.

\section{Discussion and Limitations} \label{sec:discussion}

\noindent{\bf Discussion.} Our findings show a significant potential for carbon reduction in geo-distributed LLM serving. We further observe that schedulers leave some engines idle (i.e., ready to serve but waiting for a request) while distributing requests across DCs under varying loads. For example, with the facility MILP, Llama-3.1-8B TP=1 engines remain idle for 73\%, 56\%, and 33\% of the time at low, moderate, and high loads, respectively, compared with at most 10\% under the RR policy. Depending on the load, the idle and host energy account for 28 to 81\% of the remaining carbon, over 85\% of which comes from DCs with above-median CI, motivating a load-aware engine activator that adjusts the number of active engines to the demand. We further studied tensor parallelism (TP), as it is commonly used to meet aggressive latency targets~\cite{parallelizationPaper}. With TP=2, each DC hosts half as many Llama-3.1-8B engines, each using twice the GPUs for the same batch with a sublinear speed-up, partly due to inter-GPU communication. We observe that at high load, these fewer engines saturate sooner, tightening the TPOT budget and limiting load shifts toward green DCs, so the facility MILP emits 22\% more carbon with higher SLO violations, making data-parallel deployment preferable.

\noindent{\bf Limitations.} We validate the simulator within a single node for Llama-3.1-8B only, since end-to-end data-parallel dispatch requires several engines, whereas a single Mixtral-8x7B engine at TP=4 already fills our four-GPU node. Our latency and power models are calibrated on A100 GPUs, and other accelerators would shift the absolute carbon and latency values, although our carbon--SLO findings, driven by batching and queueing dynamics, are expected to extend to them. Additionally, our workloads consist of independent single-turn requests, so while our long prompts stress the KV cache, multi-turn sessions, where prefix caching couples consecutive requests to the DC that holds their KV cache, are left for future work.

\section{Related Work}
Reducing the carbon footprint of computing has been studied along two complementary axes. Embodied carbon from hardware manufacturing is addressed through carbon-aware hardware and DC design~\cite{chasingCarbon, act, ecoServe, carbonExplorer}, while operational carbon, which depends on compute energy and grid CI, is reduced by temporally shifting delay-tolerant workloads~\cite{tempLetsWait, CAComputing}, spatially scheduling workloads across regions~\cite{casper, limitationsPaper}, or scaling resources with CI variation~\cite{carbonScaler}. For LLM serving, model-level techniques select the carbon-optimal model that meets an accuracy target~\cite{gar}, and generation-level methods shorten outputs via prompt directives~\cite{sprout}, while system-level approaches reconfigure cluster size, parallelism, and GPU frequency to reduce energy consumption~\cite{dynamoLLM, dualScale}.

As interactive LLM inference with stringent latency SLOs is not amenable to temporal shifting, existing carbon-aware LLM serving approaches exploit spatial CI variations across DCs, as summarized in Table~\ref{tab:related}. SLIT~\cite{slit} and MARLIN~\cite{marlin} schedule requests across DCs based on periodically forecasted load, using a genetic algorithm and multi-agent learning, respectively, while Green-LLM~\cite{greenllm} optimizes request flows with a linear program. FREESH~\cite{freesh} periodically provisions inference workers across DCs based on marginal emissions and load, scaling GPU frequency to meet its latency targets. GCOS~\cite{gcos} migrates GPU jobs across DCs using multi-agent reinforcement learning. Only MOER~\cite{moerRouting} operates at per-request scheduling granularity using marginal CI signals. As discussed in Section~\ref{sec:intro}, these works rely on coarse-grained decisions and batch-agnostic latency and power models without engine queueing. Meanwhile, LLM serving simulators such as Vidur~\cite{vidur} and LLMServingSim~\cite{llmservingsim, llmservingsim2} model various inference features within a single cluster, including batching, routing, disaggregated serving, KV-cache offloading, and prefix caching, yet none capture batch-dependent power, inter-DC transfers, or carbon across multiple DCs. To our knowledge, ours is the first study that jointly models continuous batching, engine queueing, inter-DC transfers, and per-DC power and carbon with hardware-calibrated latency and power models at geo-distributed scale, and evaluates carbon savings against per-token SLO attainment across various models on MLPerf benchmarks and SLOs.

\section{Conclusion} 
In this work, we present a geo-distributed LLM inference simulator, calibrated against real hardware, that models continuous batching, engine queueing, inter-DC transfers, and per-DC power and carbon. Further, we use the proposed simulator to map the carbon--SLO frontier of greedy, ILP-, and MILP-based schedulers across three models and load regimes under MLPerf SLOs. The facility MILP scheduler, which captures both batch-dependent carbon and latency within a scheduling window, offers the best carbon--SLO balance, reducing carbon emissions from 26\% to 60\% across varying loads compared to the RR policy, with an average SLO violation of 3\%. We observe that (i) the carbon savings grow with the CI spread across DCs and shrink with the load, (ii) the topology trades them against SLO violations, and (iii) SLO attainment depends on the load-update frequency. We further reveal that our findings hold under real PUE values and predicted OSLs, thereby strengthening their applicability to practical deployments. These results suggest that accounting for batch-dependent latency and power allows schedulers to reduce carbon while largely meeting per-token SLOs. As future work, we plan to extend the study to multi-turn sessions, where KV-cache locality couples consecutive placements, and to heterogeneous inference scenarios with varying models and deployment configurations across DCs.

\section*{Acknowledgment}
We used Claude (Anthropic)~\cite{claude} as an assistant for text revision, grammar checking, and code development of the simulator and analysis scripts. The research ideas, system and scheduler designs, experiments, and conclusions are the authors' own, and the authors verified all AI-assisted content.

\bibliographystyle{IEEEtran}
\bibliography{IEEEfull}

\begin{thebibliography}{10}
\providecommand{\url}[1]{#1}
\csname url@samestyle\endcsname
\providecommand{\newblock}{\relax}
\providecommand{\bibinfo}[2]{#2}
\providecommand{\BIBentrySTDinterwordspacing}{\spaceskip=0pt\relax}
\providecommand{\BIBentryALTinterwordstretchfactor}{4}
\providecommand{\BIBentryALTinterwordspacing}{\spaceskip=\fontdimen2\font plus
\BIBentryALTinterwordstretchfactor\fontdimen3\font minus \fontdimen4\font\relax}
\providecommand{\BIBforeignlanguage}[2]{{%
\expandafter\ifx\csname l@#1\endcsname\relax
\typeout{** WARNING: IEEEtran.bst: No hyphenation pattern has been}%
\typeout{** loaded for the language `#1'. Using the pattern for}%
\typeout{** the default language instead.}%
\else
\language=\csname l@#1\endcsname
\fi
#2}}
\providecommand{\BIBdecl}{\relax}
\BIBdecl

\bibitem{ai_app_1}
\BIBentryALTinterwordspacing
K.~Tomlinson, S.~Jaffe, W.~Wang, S.~Counts, and S.~Suri, ``Working with ai: Measuring the applicability of generative ai to occupations,'' 2025. [Online]. Available: \url{https://arxiv.org/abs/2507.07935}
\BIBentrySTDinterwordspacing

\bibitem{ai_app_2}
\BIBentryALTinterwordspacing
Z.~Liao, M.~Antoniak, I.~Cheong, E.~Y.-Y. Cheng, A.-H. Lee, K.~Lo, J.~C. Chang, and A.~X. Zhang, ``Llms as research tools: A large scale survey of researchers' usage and perceptions,'' 2024. [Online]. Available: \url{https://arxiv.org/abs/2411.05025}
\BIBentrySTDinterwordspacing

\bibitem{AI_IndexRep}
\BIBentryALTinterwordspacing
N.~Maslej, L.~Fattorini, R.~Perrault, Y.~Gil, V.~Parli, N.~Kariuki, E.~Capstick, A.~Reuel, E.~Brynjolfsson, J.~Etchemendy, K.~Ligett, T.~Lyons, J.~Manyika, J.~C. Niebles, Y.~Shoham, R.~Wald, T.~Walsh, A.~Hamrah, L.~Santarlasci, J.~B. Lotufo, A.~Rome, A.~Shi, and S.~Oak, ``Artificial intelligence index report 2025,'' 2025. [Online]. Available: \url{https://arxiv.org/abs/2504.07139}
\BIBentrySTDinterwordspacing

\bibitem{CoT}
\BIBentryALTinterwordspacing
J.~Wei, X.~Wang, D.~Schuurmans, M.~Bosma, B.~Ichter, F.~Xia, E.~Chi, Q.~Le, and D.~Zhou, ``Chain-of-thought prompting elicits reasoning in large language models,'' 2023. [Online]. Available: \url{https://arxiv.org/abs/2201.11903}
\BIBentrySTDinterwordspacing

\bibitem{SelfCons_CoT}
\BIBentryALTinterwordspacing
X.~Wang, J.~Wei, D.~Schuurmans, Q.~Le, E.~Chi, S.~Narang, A.~Chowdhery, and D.~Zhou, ``Self-consistency improves chain of thought reasoning in language models,'' 2023. [Online]. Available: \url{https://arxiv.org/abs/2203.11171}
\BIBentrySTDinterwordspacing

\bibitem{ReAct}
\BIBentryALTinterwordspacing
S.~Yao, J.~Zhao, D.~Yu, N.~Du, I.~Shafran, K.~Narasimhan, and Y.~Cao, ``React: Synergizing reasoning and acting in language models,'' 2023. [Online]. Available: \url{https://arxiv.org/abs/2210.03629}
\BIBentrySTDinterwordspacing

\bibitem{AutoGen}
\BIBentryALTinterwordspacing
Q.~Wu, G.~Bansal, J.~Zhang, Y.~Wu, B.~Li, E.~Zhu, L.~Jiang, X.~Zhang, S.~Zhang, J.~Liu, A.~H. Awadallah, R.~W. White, D.~Burger, and C.~Wang, ``Autogen: Enabling next-gen llm applications via multi-agent conversation,'' 2023. [Online]. Available: \url{https://arxiv.org/abs/2308.08155}
\BIBentrySTDinterwordspacing

\bibitem{CodeAgents}
\BIBentryALTinterwordspacing
X.~Du, M.~Liu, K.~Wang, H.~Wang, J.~Liu, Y.~Chen, J.~Feng, C.~Sha, X.~Peng, and Y.~Lou, ``Evaluating large language models in class-level code generation,'' in \emph{Proceedings of the IEEE/ACM 46th International Conference on Software Engineering}, ser. ICSE '24.\hskip 1em plus 0.5em minus 0.4em\relax New York, NY, USA: Association for Computing Machinery, 2024. [Online]. Available: \url{https://doi.org/10.1145/3597503.3639219}
\BIBentrySTDinterwordspacing

\bibitem{mckinsey_invest}
\BIBentryALTinterwordspacing
{McKinsey \& Company}, ``The cost of compute: A \$7 trillion race to scale data centers,'' McKinsey Quarterly, Apr. 2025, accessed: 2026-09-12. [Online]. Available: \url{https://www.mckinsey.com/industries/technology-media-and-telecommunications/our-insights/the-cost-of-compute-a-7-trillion-dollar-race-to-scale-data-centers}
\BIBentrySTDinterwordspacing

\bibitem{doe_invest1}
\BIBentryALTinterwordspacing
{U.S. Department of Energy}, ``{DOE} announces site selection for {AI} data center and energy infrastructure development on federal lands,'' Press release, Jul. 2025, accessed: 2026-09-12. [Online]. Available: \url{https://www.energy.gov/articles/doe-announces-site-selection-ai-data-center-and-energy-infrastructure-development-federal}
\BIBentrySTDinterwordspacing

\bibitem{doe_invest2}
\BIBentryALTinterwordspacing
------, ``Energy department announces new public-private partnership model, two supercomputers, to accelerate {American} dominance in science and technology,'' Press release, Oct. 2025, accessed: 2026-09-12. [Online]. Available: \url{https://www.energy.gov/articles/energy-department-announces-new-public-private-partnership-model-two-supercomputers}
\BIBentrySTDinterwordspacing

\bibitem{concern}
\BIBentryALTinterwordspacing
S.~Luccioni, Y.~Jernite, and E.~Strubell, ``Power hungry processing: Watts driving the cost of ai deployment?'' in \emph{The 2024 ACM Conference on Fairness Accountability and Transparency}, ser. FAccT ’24.\hskip 1em plus 0.5em minus 0.4em\relax ACM, June 2024, p. 85–99. [Online]. Available: \url{http://dx.doi.org/10.1145/3630106.3658542}
\BIBentrySTDinterwordspacing

\bibitem{concern_2}
\BIBentryALTinterwordspacing
C.-J. Wu, R.~Raghavendra, U.~Gupta, B.~Acun, N.~Ardalani, K.~Maeng, G.~Chang, F.~A. Behram, J.~Huang, C.~Bai, M.~Gschwind, A.~Gupta, M.~Ott, A.~Melnikov, S.~Candido, D.~Brooks, G.~Chauhan, B.~Lee, H.-H.~S. Lee, B.~Akyildiz, M.~Balandat, J.~Spisak, R.~Jain, M.~Rabbat, and K.~Hazelwood, ``Sustainable ai: Environmental implications, challenges and opportunities,'' 2022. [Online]. Available: \url{https://arxiv.org/abs/2111.00364}
\BIBentrySTDinterwordspacing

\bibitem{iea2025energyAI}
\BIBentryALTinterwordspacing
IEA, ``Energy and ai,'' 2025, licence: CC BY 4.0. Accessed: 2025-01-26. [Online]. Available: \url{https://www.iea.org/reports/energy-and-ai}
\BIBentrySTDinterwordspacing

\bibitem{ict_co2_share}
C.~Freitag, M.~Berners-Lee, K.~Widdicks, B.~Knowles, G.~S. Blair, and A.~Friday, ``The real climate and transformative impact of ict: A critique of estimates, trends, and regulations,'' \emph{Patterns}, vol.~2, no.~9, 2021.

\bibitem{reducingCarbGenAI}
\BIBentryALTinterwordspacing
A.~A. Chien, L.~Lin, H.~Nguyen, V.~Rao, T.~Sharma, and R.~Wijayawardana, ``Reducing the carbon impact of generative ai inference (today and in 2035),'' in \emph{Proceedings of the 2nd Workshop on Sustainable Computer Systems}, ser. HotCarbon '23.\hskip 1em plus 0.5em minus 0.4em\relax New York, NY, USA: Association for Computing Machinery, 2023. [Online]. Available: \url{https://doi.org/10.1145/3604930.3605705}
\BIBentrySTDinterwordspacing

\bibitem{carbonLifeCycleLLM}
\BIBentryALTinterwordspacing
P.~Jiang, C.~Sonne, W.~Li, F.~You, and S.~You, ``Preventing the immense increase in the life-cycle energy and carbon footprints of llm-powered intelligent chatbots,'' \emph{Engineering}, vol.~40, pp. 202--210, 2024. [Online]. Available: \url{https://www.sciencedirect.com/science/article/pii/S2095809924002315}
\BIBentrySTDinterwordspacing

\bibitem{Net0_AWS}
\BIBentryALTinterwordspacing
S.~K. Chaudhuri, B.~Chinnadurai, S.~Tekurkar, Vaishnav, and V.A., ``Applying carbon value modeling to achieve net-zero,'' June 2023. [Online]. Available: \url{https://aws.amazon.com/blogs/industries/applying-carbon-value-modeling-to-achieve-net-zero/}
\BIBentrySTDinterwordspacing

\bibitem{Net0_google}
\BIBentryALTinterwordspacing
B.~Gomes and K.~Brandt, ``Our 2024 environmental report,'' July 2024. [Online]. Available: \url{https://blog.google/outreach-initiatives/sustainability/2024-environmental-report/}
\BIBentrySTDinterwordspacing

\bibitem{tempLetsWait}
\BIBentryALTinterwordspacing
P.~Wiesner, I.~Behnke, D.~Scheinert, K.~Gontarska, and L.~Thamsen, ``Let’s wait awhile: how temporal workload shifting can reduce carbon emissions in the cloud,'' in \emph{Proceedings of the 22nd International Middleware Conference}, ser. Middleware ’21.\hskip 1em plus 0.5em minus 0.4em\relax ACM, Dec. 2021, p. 260–272. [Online]. Available: \url{http://dx.doi.org/10.1145/3464298.3493399}
\BIBentrySTDinterwordspacing

\bibitem{carbonCast}
\BIBentryALTinterwordspacing
D.~Maji, P.~Shenoy, and R.~K. Sitaraman, ``Carboncast: multi-day forecasting of grid carbon intensity,'' in \emph{Proceedings of the 9th ACM International Conference on Systems for Energy-Efficient Buildings, Cities, and Transportation}, ser. BuildSys '22.\hskip 1em plus 0.5em minus 0.4em\relax New York, NY, USA: Association for Computing Machinery, 2022, p. 198–207. [Online]. Available: \url{https://doi.org/10.1145/3563357.3564079}
\BIBentrySTDinterwordspacing

\bibitem{gar}
\BIBentryALTinterwordspacing
D.~Sheshanarayana, R.~S. Pal, M.~Sinha, and T.~Dasgupta, ``{GAR}: Carbon-aware routing for {LLM} inference via constrained optimization,'' 2025. [Online]. Available: \url{https://openreview.net/forum?id=wVd99lgt4j}
\BIBentrySTDinterwordspacing

\bibitem{clover}
\BIBentryALTinterwordspacing
B.~Li, S.~Samsi, V.~Gadepally, and D.~Tiwari, ``Clover: Toward sustainable ai with carbon-aware machine learning inference service,'' in \emph{Proceedings of the International Conference for High Performance Computing, Networking, Storage and Analysis}, ser. SC '23.\hskip 1em plus 0.5em minus 0.4em\relax New York, NY, USA: Association for Computing Machinery, 2023. [Online]. Available: \url{https://doi.org/10.1145/3581784.3607034}
\BIBentrySTDinterwordspacing

\bibitem{sprout}
\BIBentryALTinterwordspacing
B.~Li, Y.~Jiang, V.~Gadepally, and D.~Tiwari, ``Sprout: Green generative {AI} with carbon-efficient {LLM} inference,'' in \emph{Proceedings of the 2024 Conference on Empirical Methods in Natural Language Processing}, Y.~Al-Onaizan, M.~Bansal, and Y.-N. Chen, Eds.\hskip 1em plus 0.5em minus 0.4em\relax Miami, Florida, USA: Association for Computational Linguistics, Nov. 2024, pp. 21\,799--21\,813. [Online]. Available: \url{https://aclanthology.org/2024.emnlp-main.1215/}
\BIBentrySTDinterwordspacing

\bibitem{CompArch_PowEff}
\BIBentryALTinterwordspacing
R.~Muralidhar, R.~Borovica-Gajic, and R.~Buyya, ``Energy efficient computing systems: Architectures, abstractions and modeling to techniques and standards,'' \emph{ACM Comput. Surv.}, vol.~54, no. 11s, Sep. 2022. [Online]. Available: \url{https://doi.org/10.1145/3511094}
\BIBentrySTDinterwordspacing

\bibitem{EnEff_Sw}
\BIBentryALTinterwordspacing
R.~Pereira, M.~Couto, F.~Ribeiro, R.~Rua, J.~Cunha, J.~a.~P. Fernandes, and J.~a. Saraiva, ``Energy efficiency across programming languages: how do energy, time, and memory relate?'' in \emph{Proceedings of the 10th ACM SIGPLAN International Conference on Software Language Engineering}, ser. SLE 2017.\hskip 1em plus 0.5em minus 0.4em\relax New York, NY, USA: Association for Computing Machinery, 2017, p. 256–267. [Online]. Available: \url{https://doi.org/10.1145/3136014.3136031}
\BIBentrySTDinterwordspacing

\bibitem{EnAwareSch_DCs}
\BIBentryALTinterwordspacing
H.~Duan, C.~Chen, G.~Min, and Y.~Wu, ``Energy-aware scheduling of virtual machines in heterogeneous cloud computing systems,'' \emph{Future Generation Computer Systems}, vol.~74, pp. 142--150, 2017. [Online]. Available: \url{https://www.sciencedirect.com/science/article/pii/S0167739X16300292}
\BIBentrySTDinterwordspacing

\bibitem{throttLLM}
A.~K. Kakolyris, D.~Masouros, P.~Vavaroutsos, S.~Xydis, and D.~Soudris, ``throttll’em: Predictive gpu throttling for energy efficient llm inference serving,'' in \emph{2025 IEEE International Symposium on High Performance Computer Architecture (HPCA)}, 2025, pp. 1363--1378.

\bibitem{uServe}
\BIBentryALTinterwordspacing
H.~Qiu, W.~Mao, A.~Patke, S.~Cui, S.~Jha, C.~Wang, H.~Franke, Z.~Kalbarczyk, T.~Ba{\c s}ar, and R.~K. Iyer, ``Power-aware deep learning model serving with {$\mu$-Serve},'' in \emph{2024 USENIX Annual Technical Conference (USENIX ATC 24)}.\hskip 1em plus 0.5em minus 0.4em\relax Santa Clara, CA: USENIX Association, Jul. 2024, pp. 75--93. [Online]. Available: \url{https://www.usenix.org/conference/atc24/presentation/qiu}
\BIBentrySTDinterwordspacing

\bibitem{dynamoLLM}
\BIBentryALTinterwordspacing
J.~Stojkovic, C.~Zhang, {\'I}.~Goiri, J.~Torrellas, and E.~Choukse, ``Dynamollm: Designing llm inference clusters for performance and energy efficiency,'' in \emph{2025 IEEE International Symposium on High Performance Computer Architecture (HPCA)}.\hskip 1em plus 0.5em minus 0.4em\relax IEEE, Mar. 2025, p. 1348–1362. [Online]. Available: \url{http://dx.doi.org/10.1109/HPCA61900.2025.00102}
\BIBentrySTDinterwordspacing

\bibitem{elecMaps}
E.~Maps, ``Electricity maps,'' \url{https://app.electricitymaps.com/}, 2024, accessed: 2026-03-09.

\bibitem{gdpr}
\BIBentryALTinterwordspacing
C.~J. Hoofnagle, B.~van~der Sloot, and F.~Z. Borgesius, ``The european union general data protection regulation: what it is and what it means,'' \emph{Information \&amp; Communications Technology Law}, vol.~28, no.~1, p. 65–98, Jan. 2019. [Online]. Available: \url{http://dx.doi.org/10.1080/13600834.2019.1573501}
\BIBentrySTDinterwordspacing

\bibitem{ecovisor}
\BIBentryALTinterwordspacing
A.~Souza, N.~Bashir, J.~Murillo, W.~Hanafy, Q.~Liang, D.~Irwin, and P.~Shenoy, ``Ecovisor: A virtual energy system for carbon-efficient applications,'' in \emph{Proceedings of the 28th ACM International Conference on Architectural Support for Programming Languages and Operating Systems, Volume 2}, ser. ASPLOS 2023.\hskip 1em plus 0.5em minus 0.4em\relax New York, NY, USA: Association for Computing Machinery, 2023, p. 252–265. [Online]. Available: \url{https://doi.org/10.1145/3575693.3575709}
\BIBentrySTDinterwordspacing

\bibitem{limitationsPaper}
\BIBentryALTinterwordspacing
T.~Sukprasert, A.~Souza, N.~Bashir, D.~Irwin, and P.~Shenoy, ``On the limitations of carbon-aware temporal and spatial workload shifting in the cloud,'' in \emph{Proceedings of the Nineteenth European Conference on Computer Systems}, ser. EuroSys '24.\hskip 1em plus 0.5em minus 0.4em\relax New York, NY, USA: Association for Computing Machinery, 2024, p. 924–941. [Online]. Available: \url{https://doi.org/10.1145/3627703.3650079}
\BIBentrySTDinterwordspacing

\bibitem{carbonScaler}
\BIBentryALTinterwordspacing
W.~A. Hanafy, Q.~Liang, N.~Bashir, D.~Irwin, and P.~Shenoy, ``Carbonscaler: Leveraging cloud workload elasticity for optimizing carbon-efficiency,'' in \emph{Abstracts of the 2024 ACM SIGMETRICS/IFIP PERFORMANCE Joint International Conference on Measurement and Modeling of Computer Systems}, ser. SIGMETRICS/PERFORMANCE '24.\hskip 1em plus 0.5em minus 0.4em\relax New York, NY, USA: Association for Computing Machinery, 2024, p. 49–50. [Online]. Available: \url{https://doi.org/10.1145/3652963.3655048}
\BIBentrySTDinterwordspacing

\bibitem{casper}
\BIBentryALTinterwordspacing
A.~Souza, S.~Jasoria, B.~Chakrabarty, A.~Bridgwater, A.~Lundberg, F.~Skogh, A.~Ali-Eldin, D.~Irwin, and P.~Shenoy, ``Casper: Carbon-aware scheduling and provisioning for distributed web services,'' in \emph{Proceedings of the 14th International Green and Sustainable Computing Conference}, ser. IGSC ’23.\hskip 1em plus 0.5em minus 0.4em\relax ACM, Oct. 2023, p. 67–73. [Online]. Available: \url{http://dx.doi.org/10.1145/3634769.3634812}
\BIBentrySTDinterwordspacing

\bibitem{carbonClipper}
\BIBentryALTinterwordspacing
A.~Lechowicz, N.~Christianson, B.~Sun, N.~Bashir, M.~Hajiesmaili, A.~Wierman, and P.~Shenoy, ``Learning-augmented competitive algorithms for spatiotemporal online allocation with deadline constraints,'' 2025. [Online]. Available: \url{https://arxiv.org/abs/2408.07831}
\BIBentrySTDinterwordspacing

\bibitem{slit}
\BIBentryALTinterwordspacing
H.~Moore, S.~Qi, N.~Hogade, D.~Milojicic, C.~Bash, and S.~Pasricha, ``Sustainable carbon-aware and water-efficient llm scheduling in geo-distributed cloud datacenters,'' in \emph{Proceedings of the Great Lakes Symposium on VLSI 2025}, ser. GLSVLSI '25.\hskip 1em plus 0.5em minus 0.4em\relax New York, NY, USA: Association for Computing Machinery, 2025, p. 929–934. [Online]. Available: \url{https://doi.org/10.1145/3716368.3735301}
\BIBentrySTDinterwordspacing

\bibitem{greenllm}
\BIBentryALTinterwordspacing
J.~Cheng and D.~T. Nguyen, ``Green-llm: Optimal workload allocation for environmentally-aware distributed inference,'' 2026. [Online]. Available: \url{https://arxiv.org/abs/2507.09942}
\BIBentrySTDinterwordspacing

\bibitem{freesh}
\BIBentryALTinterwordspacing
X.~He, Z.~Fang, J.~Lian, D.~H.~K. Tsang, B.~Zhang, and Y.~Chen, ``Freesh: Fair, resource- and energy-efficient scheduling for llm serving on heterogeneous gpus,'' 2025. [Online]. Available: \url{https://arxiv.org/abs/2511.00807}
\BIBentrySTDinterwordspacing

\bibitem{gcos}
\BIBentryALTinterwordspacing
H.~Liu, X.~Hu, R.~Wang, J.~Hao, Q.~Wu, and H.~Zhang, ``Green scheduling for llm workloads with model and data reuse across geo-distributed data centers,'' \emph{Digital Communications and Networks}, vol.~12, no.~2, pp. 236--251, 2026. [Online]. Available: \url{https://www.sciencedirect.com/science/article/pii/S2352864825001828}
\BIBentrySTDinterwordspacing

\bibitem{marlin}
\BIBentryALTinterwordspacing
H.~Moore, S.~Qi, D.~Milojicic, C.~Bash, and S.~Pasricha, ``Marlin: Multi-agent game-theoretic reinforcement learning for sustainable llm inference in cloud datacenters,'' in \emph{Proceedings of the 16th ACM International Green and Sustainable Computing Conference}, ser. IGSC '26.\hskip 1em plus 0.5em minus 0.4em\relax New York, NY, USA: Association for Computing Machinery, 2026, p. 28–35. [Online]. Available: \url{https://doi.org/10.1145/3797248.3815404}
\BIBentrySTDinterwordspacing

\bibitem{moerRouting}
\BIBentryALTinterwordspacing
A.~Bernhard and A.~B. Yardimci, ``Routing llm inference to the cleanest grid in real time,'' 2026. [Online]. Available: \url{https://arxiv.org/abs/2608.06188}
\BIBentrySTDinterwordspacing

\bibitem{orca}
\BIBentryALTinterwordspacing
G.-I. Yu, J.~S. Jeong, G.-W. Kim, S.~Kim, and B.-G. Chun, ``Orca: A distributed serving system for {Transformer-Based} generative models,'' in \emph{16th USENIX Symposium on Operating Systems Design and Implementation (OSDI 22)}.\hskip 1em plus 0.5em minus 0.4em\relax Carlsbad, CA: USENIX Association, Jul. 2022, pp. 521--538. [Online]. Available: \url{https://www.usenix.org/conference/osdi22/presentation/yu}
\BIBentrySTDinterwordspacing

\bibitem{splitWise}
\BIBentryALTinterwordspacing
P.~Patel, E.~Choukse, C.~Zhang, A.~Shah, Íñigo Goiri, S.~Maleki, and R.~Bianchini, ``Splitwise: Efficient generative llm inference using phase splitting,'' 2024. [Online]. Available: \url{https://arxiv.org/abs/2311.18677}
\BIBentrySTDinterwordspacing

\bibitem{mlperf}
\BIBentryALTinterwordspacing
V.~J. Reddi, C.~Cheng, D.~Kanter, P.~Mattson, G.~Schmuelling, C.-J. Wu, B.~Anderson, M.~Breughe, M.~Charlebois, W.~Chou, R.~Chukka, C.~Coleman, S.~Davis, P.~Deng, G.~Diamos, J.~Duke, D.~Fick, J.~S. Gardner, I.~Hubara, S.~Idgunji, T.~B. Jablin, J.~Jiao, T.~S. John, P.~Kanwar, D.~Lee, J.~Liao, A.~Lokhmotov, F.~Massa, P.~Meng, P.~Micikevicius, C.~Osborne, G.~Pekhimenko, A.~T.~R. Rajan, D.~Sequeira, A.~Sirasao, F.~Sun, H.~Tang, M.~Thomson, F.~Wei, E.~Wu, L.~Xu, K.~Yamada, B.~Yu, G.~Yuan, A.~Zhong, P.~Zhang, and Y.~Zhou, ``Mlperf inference benchmark,'' in \emph{Proceedings of the ACM/IEEE 47th Annual International Symposium on Computer Architecture}, ser. ISCA '20.\hskip 1em plus 0.5em minus 0.4em\relax IEEE Press, 2020, p. 446–459. [Online]. Available: \url{https://doi.org/10.1109/ISCA45697.2020.00045}
\BIBentrySTDinterwordspacing

\bibitem{cloudping}
{Matt Adorjan}, ``{CloudPing}: {AWS} inter-region latency monitoring,'' \url{https://www.cloudping.co/}, 2026, one-year p50 round-trip latency between AWS regions. Accessed: 2026-09-27.

\bibitem{ipcc_ci}
S.~Schloemer, T.~Bruckner, L.~Fulton, E.~Hertwich, A.~McKinnon, D.~Perczyk, J.~Roy, R.~Schaeffer, R.~Sims, P.~Smith, and R.~Wiser, \emph{\BIBforeignlanguage{English}{Annex III: Technology-specific cost and performance parameters}}.\hskip 1em plus 0.5em minus 0.4em\relax United Kingdom: Cambridge University Press, 2014, pp. 1329--1356, this annex should be cited as: Schl{\"o}mer S., T. Bruckner, L. Fulton, E. Hertwich, A. McKinnon, D. Perczyk, J. Roy, R. Schaeffer, R. Sims, P. Smith, and R. Wiser, 2014: Annex III: Technology-specific cost and performance parameters. In: Climate Change 2014: Mitigation of Climate Change. Contribution of Working Group III to the Fifth Assessment Report of the Intergovernmental Panel on Climate Change [Edenhofer, O., R. Pichs-Madruga, Y. Sokona, E. Farahani, S. Kadner, K. Seyboth, A. Adler, I. Baum, S. Brunner, P. Eickemeier, B. Kriemann, J. Savolainen, S. Schl{\"o}mer, C. von Stechow, T. Zwickel and J.C. Minx (eds.)]. Cambridge University Press, Cambridge, United Kingdom and New
  York, NY, USA.

\bibitem{GHGScopes}
\BIBentryALTinterwordspacing
G.~G. Protocol, ``Greenhouse gas protocol,'' \emph{Sector Toolsets for Iron and Steel-Guidance Document}, pp. 1--12, 2011. [Online]. Available: \url{https://www.itf-oecd.org/sites/default/files/1.2._ghg_protocol.pdf}
\BIBentrySTDinterwordspacing

\bibitem{chasingCarbon}
U.~Gupta, Y.~G. Kim, S.~Lee, J.~Tse, H.-H.~S. Lee, G.-Y. Wei, D.~Brooks, and C.-J. Wu, ``Chasing carbon: The elusive environmental footprint of computing,'' in \emph{2021 IEEE International Symposium on High-Performance Computer Architecture (HPCA)}, 2021, pp. 854--867.

\bibitem{distServe}
\BIBentryALTinterwordspacing
Y.~Zhong, S.~Liu, J.~Chen, J.~Hu, Y.~Zhu, X.~Liu, X.~Jin, and H.~Zhang, ``Distserve: Disaggregating prefill and decoding for goodput-optimized large language model serving,'' 2024. [Online]. Available: \url{https://arxiv.org/abs/2401.09670}
\BIBentrySTDinterwordspacing

\bibitem{sarathiServe}
A.~Agrawal, N.~Kedia, A.~Panwar, J.~Mohan, N.~Kwatra, B.~S. Gulavani, A.~Tumanov, and R.~Ramjee, ``Taming throughput-latency tradeoff in llm inference with sarathi-serve,'' in \emph{Proceedings of the 18th USENIX Conference on Operating Systems Design and Implementation}, ser. OSDI'24.\hskip 1em plus 0.5em minus 0.4em\relax USA: USENIX Association, 2024.

\bibitem{AWS}
{AWS}, ``Amazon bedrock now supports global cross-region inference for anthropic claude sonnet 4,'' \url{https://aws.amazon.com/about-aws/whats-new/2025/09/amazon-bedrock-global-cross-region-inference-anthropic-claude-sonnet-4/}, September 2025, accessed: 2026-04-27.

\bibitem{skywalker}
\BIBentryALTinterwordspacing
T.~Xia, Z.~Mao, J.~Kerney, E.~J. Jackson, Z.~Li, J.~Xing, S.~Shenker, and I.~Stoica, ``Skywalker: A locality-aware cross-region load balancer for llm inference,'' 2025. [Online]. Available: \url{https://arxiv.org/abs/2505.24095}
\BIBentrySTDinterwordspacing

\bibitem{central_Spanner}
\BIBentryALTinterwordspacing
J.~C. Corbett, J.~Dean, M.~Epstein, A.~Fikes, C.~Frost, J.~J. Furman, S.~Ghemawat, A.~Gubarev, C.~Heiser, P.~Hochschild, W.~Hsieh, S.~Kanthak, E.~Kogan, H.~Li, A.~Lloyd, S.~Melnik, D.~Mwaura, D.~Nagle, S.~Quinlan, R.~Rao, L.~Rolig, Y.~Saito, M.~Szymaniak, C.~Taylor, R.~Wang, and D.~Woodford, ``Spanner: Google’s globally distributed database,'' \emph{ACM Trans. Comput. Syst.}, vol.~31, no.~3, Aug. 2013. [Online]. Available: \url{https://doi.org/10.1145/2491245}
\BIBentrySTDinterwordspacing

\bibitem{dynamo_amazon}
\BIBentryALTinterwordspacing
G.~DeCandia, D.~Hastorun, M.~Jampani, G.~Kakulapati, A.~Lakshman, A.~Pilchin, S.~Sivasubramanian, P.~Vosshall, and W.~Vogels, ``Dynamo: amazon's highly available key-value store,'' in \emph{Proceedings of Twenty-First ACM SIGOPS Symposium on Operating Systems Principles}, ser. SOSP '07.\hskip 1em plus 0.5em minus 0.4em\relax New York, NY, USA: Association for Computing Machinery, 2007, p. 205–220. [Online]. Available: \url{https://doi.org/10.1145/1294261.1294281}
\BIBentrySTDinterwordspacing

\bibitem{hierarchical_apollo}
\BIBentryALTinterwordspacing
E.~Boutin, J.~Ekanayake, W.~Lin, B.~Shi, J.~Zhou, Z.~Qian, M.~Wu, and L.~Zhou, ``Apollo: Scalable and coordinated scheduling for {Cloud-Scale} computing,'' in \emph{11th USENIX Symposium on Operating Systems Design and Implementation (OSDI 14)}.\hskip 1em plus 0.5em minus 0.4em\relax Broomfield, CO: USENIX Association, Oct. 2014, pp. 285--300. [Online]. Available: \url{https://www.usenix.org/conference/osdi14/technical-sessions/presentation/boutin}
\BIBentrySTDinterwordspacing

\bibitem{carbonAwareLoadBal_1}
\BIBentryALTinterwordspacing
Z.~Zhou, F.~Liu, Y.~Xu, R.~Zou, H.~Xu, J.~C.~S. Lui, and H.~Jin, ``Carbon-aware load balancing for geo-distributed cloud services,'' in \emph{Proceedings of the 2013 IEEE 21st International Symposium on Modelling, Analysis \& Simulation of Computer and Telecommunication Systems}, ser. MASCOTS '13.\hskip 1em plus 0.5em minus 0.4em\relax USA: IEEE Computer Society, 2013, p. 232–241. [Online]. Available: \url{https://doi.org/10.1109/MASCOTS.2013.31}
\BIBentrySTDinterwordspacing

\bibitem{Llumnix}
B.~Sun, Z.~Huang, H.~Zhao, W.~Xiao, X.~Zhang, Y.~Li, and W.~Lin, ``Llumnix: dynamic scheduling for large language model serving,'' in \emph{Proceedings of the 18th USENIX Conference on Operating Systems Design and Implementation}, ser. OSDI'24.\hskip 1em plus 0.5em minus 0.4em\relax USA: USENIX Association, 2024.

\bibitem{parallelizationPaper}
\BIBentryALTinterwordspacing
B.~Topcu, M.~O. Cim, P.~Palangappa, M.~Arunachalam, and M.~T. Kandemir, ``Parallelization strategies for dense llm deployment: Navigating through application-specific tradeoffs and bottlenecks,'' 2026. [Online]. Available: \url{https://arxiv.org/abs/2603.05692}
\BIBentrySTDinterwordspacing

\bibitem{moe_cost_break}
\BIBentryALTinterwordspacing
H.~Huang, N.~Ardalani, A.~Sun, L.~Ke, H.-H.~S. Lee, S.~Bhosale, C.-J. Wu, and B.~Lee, ``Toward efficient inference for mixture of experts,'' in \emph{Advances in Neural Information Processing Systems}, A.~Globerson, L.~Mackey, D.~Belgrave, A.~Fan, U.~Paquet, J.~Tomczak, and C.~Zhang, Eds., vol.~37.\hskip 1em plus 0.5em minus 0.4em\relax Curran Associates, Inc., 2024, pp. 84\,033--84\,059. [Online]. Available: \url{https://proceedings.neurips.cc/paper_files/paper/2024/file/98bf3b8505c611ac21055dd9d355c66e-Paper-Conference.pdf}
\BIBentrySTDinterwordspacing

\bibitem{llama3}
\BIBentryALTinterwordspacing
{Llama Team}, ``The llama 3 herd of models,'' \emph{CoRR}, vol. abs/2407.21783, 2024. [Online]. Available: \url{https://doi.org/10.48550/arXiv.2407.21783}
\BIBentrySTDinterwordspacing

\bibitem{qwen}
\BIBentryALTinterwordspacing
Qwen, :, A.~Yang, B.~Yang, B.~Zhang, B.~Hui, B.~Zheng, B.~Yu, C.~Li, D.~Liu, F.~Huang, H.~Wei, H.~Lin, J.~Yang, J.~Tu, J.~Zhang, J.~Yang, J.~Yang, J.~Zhou, J.~Lin, K.~Dang, K.~Lu, K.~Bao, K.~Yang, L.~Yu, M.~Li, M.~Xue, P.~Zhang, Q.~Zhu, R.~Men, R.~Lin, T.~Li, T.~Tang, T.~Xia, X.~Ren, X.~Ren, Y.~Fan, Y.~Su, Y.~Zhang, Y.~Wan, Y.~Liu, Z.~Cui, Z.~Zhang, and Z.~Qiu, ``Qwen2.5 technical report,'' 2025. [Online]. Available: \url{https://arxiv.org/abs/2412.15115}
\BIBentrySTDinterwordspacing

\bibitem{mixtral}
\BIBentryALTinterwordspacing
A.~Q. Jiang, A.~Sablayrolles, A.~Roux, A.~Mensch, B.~Savary, C.~Bamford, D.~S. Chaplot, D.~de~las Casas, E.~B. Hanna, F.~Bressand, G.~Lengyel, G.~Bour, G.~Lample, L.~R. Lavaud, L.~Saulnier, M.-A. Lachaux, P.~Stock, S.~Subramanian, S.~Yang, S.~Antoniak, T.~L. Scao, T.~Gervet, T.~Lavril, T.~Wang, T.~Lacroix, and W.~E. Sayed, ``Mixtral of experts,'' 2024. [Online]. Available: \url{https://arxiv.org/abs/2401.04088}
\BIBentrySTDinterwordspacing

\bibitem{humanEval}
\BIBentryALTinterwordspacing
M.~Chen, J.~Tworek, H.~Jun, Q.~Yuan, H.~P. de~Oliveira~Pinto, J.~Kaplan, H.~Edwards, Y.~Burda, N.~Joseph, G.~Brockman, A.~Ray, R.~Puri, G.~Krueger, M.~Petrov, H.~Khlaaf, G.~Sastry, P.~Mishkin, B.~Chan, S.~Gray, N.~Ryder, M.~Pavlov, A.~Power, L.~Kaiser, M.~Bavarian, C.~Winter, P.~Tillet, F.~P. Such, D.~Cummings, M.~Plappert, F.~Chantzis, E.~Barnes, A.~Herbert-Voss, W.~H. Guss, A.~Nichol, A.~Paino, N.~Tezak, J.~Tang, I.~Babuschkin, S.~Balaji, S.~Jain, W.~Saunders, C.~Hesse, A.~N. Carr, J.~Leike, J.~Achiam, V.~Misra, E.~Morikawa, A.~Radford, M.~Knight, M.~Brundage, M.~Murati, K.~Mayer, P.~Welinder, B.~McGrew, D.~Amodei, S.~McCandlish, I.~Sutskever, and W.~Zaremba, ``Evaluating large language models trained on code,'' 2021. [Online]. Available: \url{https://arxiv.org/abs/2107.03374}
\BIBentrySTDinterwordspacing

\bibitem{mbpp}
\BIBentryALTinterwordspacing
J.~Austin, A.~Odena, M.~Nye, M.~Bosma, H.~Michalewski, D.~Dohan, E.~Jiang, C.~Cai, M.~Terry, Q.~Le, and C.~Sutton, ``Program synthesis with large language models,'' 2021. [Online]. Available: \url{https://arxiv.org/abs/2108.07732}
\BIBentrySTDinterwordspacing

\bibitem{govReport}
\BIBentryALTinterwordspacing
L.~Huang, S.~Cao, N.~Parulian, H.~Ji, and L.~Wang, ``Efficient attentions for long document summarization,'' 2021. [Online]. Available: \url{https://arxiv.org/abs/2104.02112}
\BIBentrySTDinterwordspacing

\bibitem{a100}
{NVIDIA Corp.}, ``Nvidia a100 tensor core gpu datasheet,'' \url{https://www.nvidia.com/content/dam/en-zz/Solutions/Data-Center/a100/pdf/nvidia-a100-datasheet-us-nvidia-1758950-r4-web.pdf}, 2022, accessed: 2026-04-30.

\bibitem{vllm_engine_args}
vLLM, ``vllm engine arguments,'' \url{https://docs.vllm.ai/en/v0.6.0/models/engine_args.html}, 2024, vLLM v0.6.0 documentation, Accessed 2026-04-28.

\bibitem{highs}
\BIBentryALTinterwordspacing
Q.~Huangfu and J.~A.~J. Hall, ``Parallelizing the dual revised simplex method,'' 2015. [Online]. Available: \url{https://arxiv.org/abs/1503.01889}
\BIBentrySTDinterwordspacing

\bibitem{google_pue}
Google, ``Growing the internet while reducing energy consumption,'' \url{https://datacenters.google/efficiency/}, 2025, accessed: Oct. 5, 2026.

\bibitem{realPUE}
\BIBentryALTinterwordspacing
G.~Attenni and N.~Bartolini, ``Environmentally-conscious cloud orchestration considering geo-distributed data centers,'' 2025. [Online]. Available: \url{https://arxiv.org/abs/2507.11563}
\BIBentrySTDinterwordspacing

\bibitem{mlperfInfer}
{MLCommons}, ``{MLPerf} inference rules,'' \url{https://github.com/mlcommons/inference_policies/blob/master/inference_rules.adoc}, 2025, benchmarks section, Server scenario latency constraints. Accessed: 2026-09-28.

\bibitem{cnn_daily}
K.~M. Hermann, T.~Ko{\v c}isk{\'y}, E.~Grefenstette, L.~Espeholt, W.~Kay, M.~Suleyman, and P.~Blunsom, ``Teaching machines to read and comprehend,'' in \emph{Proceedings of the 29th International Conference on Neural Information Processing Systems - Volume 1}, ser. NIPS'15.\hskip 1em plus 0.5em minus 0.4em\relax Cambridge, MA, USA: MIT Press, 2015, p. 1693–1701.

\bibitem{OpenOrca}
W.~Lian, B.~Goodson, E.~Pentland, A.~Cook, C.~Vong, and "Teknium", ``Openorca: An open dataset of gpt augmented flan reasoning traces,'' \url{https://huggingface.co/datasets/Open-Orca/OpenOrca}, 2023.

\bibitem{OSLpred}
\BIBentryALTinterwordspacing
H.~Qiu, W.~Mao, A.~Patke, S.~Cui, S.~Jha, C.~Wang, H.~Franke, Z.~T. Kalbarczyk, T.~Başar, and R.~K. Iyer, ``Efficient interactive llm serving with proxy model-based sequence length prediction,'' 2024. [Online]. Available: \url{https://arxiv.org/abs/2404.08509}
\BIBentrySTDinterwordspacing

\bibitem{gsm8k}
\BIBentryALTinterwordspacing
K.~Cobbe, V.~Kosaraju, M.~Bavarian, M.~Chen, H.~Jun, L.~Kaiser, M.~Plappert, J.~Tworek, J.~Hilton, R.~Nakano, C.~Hesse, and J.~Schulman, ``Training verifiers to solve math word problems,'' 2021. [Online]. Available: \url{https://arxiv.org/abs/2110.14168}
\BIBentrySTDinterwordspacing

\bibitem{act}
\BIBentryALTinterwordspacing
U.~Gupta, M.~Elgamal, G.~Hills, G.-Y. Wei, H.-H.~S. Lee, D.~Brooks, and C.-J. Wu, ``Act: designing sustainable computer systems with an architectural carbon modeling tool,'' in \emph{Proceedings of the 49th Annual International Symposium on Computer Architecture}, ser. ISCA '22.\hskip 1em plus 0.5em minus 0.4em\relax New York, NY, USA: Association for Computing Machinery, 2022, p. 784–799. [Online]. Available: \url{https://doi.org/10.1145/3470496.3527408}
\BIBentrySTDinterwordspacing

\bibitem{ecoServe}
\BIBentryALTinterwordspacing
Y.~Li, Z.~Hu, E.~Choukse, R.~Fonseca, G.~E. Suh, and U.~Gupta, ``Ecoserve: Designing carbon-aware ai inference systems,'' 2025. [Online]. Available: \url{https://arxiv.org/abs/2502.05043}
\BIBentrySTDinterwordspacing

\bibitem{carbonExplorer}
\BIBentryALTinterwordspacing
B.~Acun, B.~Lee, F.~Kazhamiaka, K.~Maeng, U.~Gupta, M.~Chakkaravarthy, D.~Brooks, and C.-J. Wu, ``Carbon explorer: A holistic framework for designing carbon aware datacenters,'' in \emph{Proceedings of the 28th ACM International Conference on Architectural Support for Programming Languages and Operating Systems, Volume 2}, ser. ASPLOS 2023.\hskip 1em plus 0.5em minus 0.4em\relax New York, NY, USA: Association for Computing Machinery, 2023, p. 118–132. [Online]. Available: \url{https://doi.org/10.1145/3575693.3575754}
\BIBentrySTDinterwordspacing

\bibitem{CAComputing}
A.~Radovanović, R.~Koningstein, I.~Schneider, B.~Chen, A.~Duarte, B.~Roy, D.~Xiao, M.~Haridasan, P.~Hung, N.~Care, S.~Talukdar, E.~Mullen, K.~Smith, M.~Cottman, and W.~Cirne, ``Carbon-aware computing for datacenters,'' \emph{IEEE Transactions on Power Systems}, vol.~38, no.~2, pp. 1270--1280, 2023.

\bibitem{dualScale}
\BIBentryALTinterwordspacing
O.~Basit, Y.~Liu, Z.~J. Kong, and Y.~C. Hu, ``Dualscale: Energy-efficient disaggregated llm serving via phase-aware placement and dvfs,'' 2026. [Online]. Available: \url{https://arxiv.org/abs/2602.18755}
\BIBentrySTDinterwordspacing

\bibitem{vidur}
\BIBentryALTinterwordspacing
A.~Agrawal, N.~Kedia, J.~Mohan, A.~Panwar, N.~Kwatra, B.~Gulavani, R.~Ramjee, and A.~Tumanov, ``Vidur: A large-scale simulation framework for llm inference,'' 2024. [Online]. Available: \url{https://arxiv.org/abs/2405.05465}
\BIBentrySTDinterwordspacing

\bibitem{llmservingsim}
\BIBentryALTinterwordspacing
J.~Cho, M.~Kim, H.~Choi, G.~Heo, and J.~Park, ``Llmservingsim: A hw/sw co-simulation infrastructure for llm inference serving at scale,'' in \emph{2024 IEEE International Symposium on Workload Characterization (IISWC)}.\hskip 1em plus 0.5em minus 0.4em\relax IEEE, Sep. 2024, p. 15–29. [Online]. Available: \url{http://dx.doi.org/10.1109/IISWC63097.2024.00012}
\BIBentrySTDinterwordspacing

\bibitem{llmservingsim2}
\BIBentryALTinterwordspacing
J.~Cho, H.~Choi, G.~Heo, and J.~Park, ``Llmservingsim 2.0: A unified simulator for heterogeneous and disaggregated llm serving infrastructure,'' 2026. [Online]. Available: \url{https://arxiv.org/abs/2602.23036}
\BIBentrySTDinterwordspacing

\bibitem{claude}
Anthropic, ``Claude opus 5.5,'' \url{https://www.anthropic.com/claude-opus-5-5}, 2026, large language model; accessed October 2026.

\end{thebibliography}

\end{document}